\documentclass[final]{ws-procs9x6}
\usepackage{graphics}
\usepackage{axodraw}
\usepackage{natbib}

\newcommand{ \slashchar }[1]{\setbox0=\hbox{$#1$}   
   \dimen0=\wd0                                     
   \setbox1=\hbox{/} \dimen1=\wd1                   
   \ifdim\dimen0>\dimen1                            
      \rlap{\hbox to \dimen0{\hfil/\hfil}}          
      #1                                            
   \else                                            
      \rlap{\hbox to \dimen1{\hfil$#1$\hfil}}       
      /                                             
   \fi}                                             %

\newcommand{\Mpl}{M_\mathrm{Pl}}
\newcommand{\ra}{\rightarrow}
\newcommand{\Mstar}{M_\star}
\newcommand{\MRS}{M_\mathrm{RS}}
\newcommand{\gsim}{\gtrsim}
\newcommand{\lsim}{\lesssim}

\newcommand{\B}{B^{(1)}}

\bibpunct{[}{]}{,}{n}{,}{,}

\begin{document}

\def\draftnote{}
\def\trimmarks{}
\setlength{\topmargin}{-0.01cm}

\pagestyle{plain}

\title{TASI 2004 Lectures on the \\ Phenomenology of Extra Dimensions}

\author{GRAHAM D. KRIBS}

\address{Department of Physics and Institute of Theoretical Science, \\
University of Oregon, Eugene, OR 97403 \\[3mm]
\textnormal{\texttt{kribs@uoregon.edu}}}

\maketitle

\abstracts{The phenomenology of large, warped, and universal 
extra dimensions is reviewed.  Characteristic signals are emphasized
rather than an extensive survey.  This is the writeup of lectures given at 
the Theoretical Advanced Study Institute in 2004.}

\section{Introduction}

The most exciting development in physics beyond the
Standard Model in the past ten years is the phenomenology of
extra dimensions.  A cursory glance at the SLAC Spires 
``all time high'' citation count confirms this crude statement.
As of the close of the summer of 2005, the papers on 
large \cite{Arkani-Hamed:1998rs}, 
warped \cite{Randall:1999ee}, 
and universal \cite{Appelquist:2000nn}
extra dimensions have garnered nearly 4000 
citations\footnote{Nearly 4000 papers have cited one or more 
these three papers.} among them.  This also shows that the literature
on extra dimensions is immense.  As a consequence, I will utterly
fail at being able to provide a complete reference list, 
and hence I apologize in advance for omissions.

Extra dimensions have been around for a long time.
Kaluza and Klein postulated a fifth dimension to 
unify electromagnetism with gravity \cite{Kaluza:1921tu}.  
A closer look at
this old idea reveals both its promise and problems.
Imagine a Universe with five-dimensional (5-d) gravity, 
in which one dimension is compactified on a circle
with circumference $L$.
The Einstein-Hilbert action is
\begin{equation}
S = \int d^5 x \sqrt{-g^{(5)}} \, \Mstar^3 R^{(5)} 
\end{equation}
where $g_{M N}^{(5)}$ is the metric and $R^{(5)}$ is the Ricci scalar 
for the five-dimensional spacetime, respectively.
Expanding the metric about flat spacetime, 
$g_{M N} = \eta_{M N} + h_{M N}/(2 \Mstar^{3/2})$, 
the five-dimensional graviton 
$h_{M N}$ contains five physical
components that are decomposed
on $R_4 \otimes S^1$ at the massless level as
\begin{equation}
\left( \begin{array}{cc}
       h_{\mu\nu} & A_{\mu 5} \\
       A_{5\nu}   & \phi      
       \end{array} \right)
\end{equation}
where $h_{\mu\nu}$ is the four-dimensional (4-d) graviton, 
$A_{\mu 5}$ is a massless vector field, and
$\phi$ is a massless scalar field.
The action reduces to
\begin{equation}
\int d^4 x (M_\star^3 L) R^{(4)} - \frac{1}{4} F_{\mu\nu} F^{\mu\nu} 
+ \frac{1}{2} \partial_\mu \phi \partial^\mu \phi
\end{equation}
comprising four-dimensional gravity plus a gauge field with 
coupling strength $g^2 = (M_* L)^{-1}$, as well as a massless 
scalar field (with only gravitational couplings).

The remarkable finding that gauge theory could arise
from a higher dimensional spacetime suitably compactified
has been a tantalizing hint of how to unify gravity with
the other gauge forces.  As it stands, however, the original
Kaluza-Klein proposal suffers from three problems: \\
(1) there is a gravitationally coupled scalar field $\phi$; \\
(2) the gauge field strength is order one only when $L^{-1} \sim M_*$; \\
(3) fermions are not chiral in five dimensions, leading to 
fermion ``doubling'' at the massless level.  \\
Orbifolding the compactified spacetime on $S^1/Z_2$ doesn't help, 
since the same operation that projects out half of the fermions 
also projects out the massless gauge field.

Where the original hope of Kaluza-Klein's idea fails, string
theory takes over, and I refer you to other TASI lectures
and books to give you the past and present scoop on string theory
(for starters, try Ref.~\cite{Zwiebach:2004tj}).
These lectures, instead, concentrate on what the world is
like if some of or all of the fields we know and love live in
extra dimensions.  There is some overlap between these lectures
and those of Sundrum \cite{Sundrum:2005jf} 
and Cs\'aki (with Hubisz and Meade) \cite{Csaki:2005vy}, however, 
I believe you will find that my perspective on this 
(and the direction given to me by the TASI 2004 organizers) 
is somewhat different.  Hopefully it is useful!

There is one issue that I think is useful to dispense with 
right away, namely:  Why study extra dimensions?  
In light of deconstruction \cite{Arkani-Hamed:2001ca,Hill:2000mu},
one is tempted to believe everything can be studied from a 
purely four-dimensional view.  This is certainly true of gauge theory.  
Does this mean
we should discard extra dimensions and just consider product gauge theories?
Here it useful to consider the point of view of 
Hill, Pokorski, Wang \cite{Hill:2000mu},
in which they sought an effective theory of the Kaluza-Klein modes
of an extra dimension.  They emphasized that imposing a lattice
cutoff on the extra dimensional space was tantamount to writing
a product gauge theory with particular relationships among all 
of the couplings and masses.  Perhaps an analogy to gauge theory
is useful here.  Imagine that you know nothing of gauge theory
and just go out and measure couplings of fermions to gluons 
and gluons to themselves (3-point and 4-point couplings).
Gradually, through careful measurement you would find that 
the couplings are all related, up to certain overall
constants (later identified as group theory factors dependent
on the representation of the fermions).  These relationships would be 
curious, but certainly would not prevent you from writing down
the low energy effective theory of the couplings of these particles.
Eventually, once the couplings are established to be the same 
up to some experimental accuracy, you would discard the effective
theory of totally separate couplings and instead just write down
the QCD Lagrangian.  Analogously, once the couplings and masses
of Kaluza-Klein modes are measured to sufficient accuracy, 
one will likely cease to characterize this as ``a product gauge theory
with relationships among the couplings'' and instead simply 
begin saying one has ``discovered an extra dimension''.

All of this is true for both gauge theory and gravity.  However,
deconstructing gravity has proved more elusive, due to 
various issues of strong coupling that appear inevitable. 
I refer you to several papers \cite{deconstructinggravity}
for discussions of this fascinating topic.  Suffice to say, 
it is much more straightforward to understand the low energy 
effective theory of a compactified ``physical'' extra dimension 
rather than an extra dimension built out of multiple general
coordinate invariances.  Since the focus of these lectures is
on phenomenology, this view is efficient, simple, and prudent.

\section{Large Extra Dimensions}

\begin{figure}
\begin{center}
\begin{picture}(300,200)

  \Line(  50,  50 )( 150, 100 )
  \Line( 150, 100 )( 250,  50 )
  \Line( 150,   0 )( 250,  50 )
  \Line( 150,   0 )(  50,  50 )
  \Text( 260,  20 )[c]{4-d spacetime}
  \DashLine( 70, 40  )( 170, 90 ){4}
  \DashLine( 90, 30  )( 190, 80 ){4}
  \DashLine( 110, 20  )( 210, 70 ){4}
  \DashLine( 130, 10  )( 230, 60 ){4}
  \DashLine( 70, 60  )( 170, 10 ){4}
  \DashLine( 90, 70  )( 190, 20 ){4}
  \DashLine( 110, 80  )( 210, 30 ){4}
  \DashLine( 130, 90  )( 230, 40 ){4}
  \CArc( 150, 150 )(50, 0, 360)
  \qbezier( 150, 200 )( 100, 150 )( 150, 100 )
  \qbezier( 150, 200 )( 200, 150 )( 150, 100 )
  \Text( 260, 157 )[c]{extra compactified}
  \Text( 260, 143 )[c]{dimensions}
  \LongArrow( 95, 170 )( 105, 185 )
  \Text( 90, 165 )[c]{$y_i$}
  \LongArrow( 120, 160 )( 125, 175 )
  \Text( 115, 155 )[c]{$y_j$}
\end{picture}
\end{center}
\caption{Sketch of the large extra dimension ADD model worldview.}
\label{ADD-sketch-fig}
\end{figure}
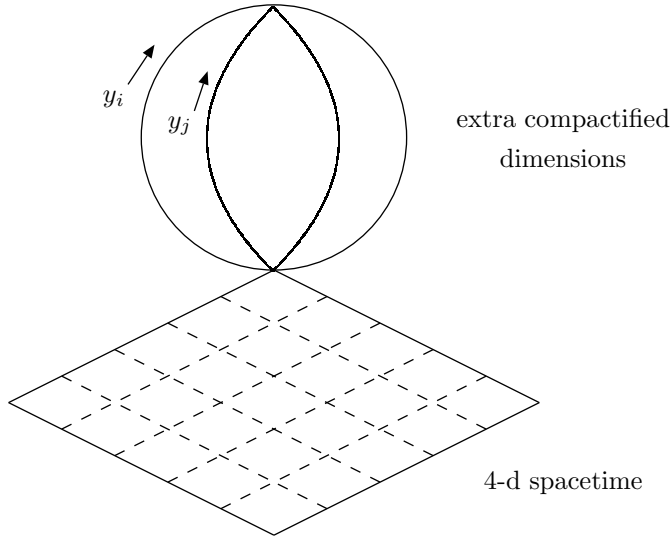

The renaissance of extra dimensions began with the 
Arkani-Hamed, Dimopoulos, and Dvali (ADD) proposal 
\cite{Arkani-Hamed:1998rs,Arkani-Hamed:1998nn} to lower 
the scale of quantum gravity to a TeV by localizing the SM to 
a 3+1 dimensional surface or ``brane'' in a higher dimensional spacetime.
The extra dimensions are compactified into a large volume that
effectively dilutes the strength of gravity from the fundamental 
scale (the TeV scale) to the Planck scale.  A sketch of the
setup is shown in Fig.~\ref{ADD-sketch-fig}.

The idea that the quantum gravity scale could be lowered
while the SM remain on a brane was motivated by earlier results in 
string theory.  In particular, it was realized in string theory
that the quantum gravity scale could be lowered from the
Planck scale to the GUT scale \cite{stringextraD}.
Others also pursued 
extra dimensions opening up between the TeV scale 
to the GUT scale \cite{otherearly}.
In this section, however, I will concentrate solely on
the ADD model and discuss several of its important 
phenomenological implications.

First, let's be explicit about the assumptions.  
The ADD model consists of 
\begin{itemize}
\item $n$ extra dimensions, each compactified with radius $r$
      (taken to be the same size for each dimension) 
      on a torus with volume $V_n = (2 \pi r)^n$.
\item All SM fields (matter, Higgs, gauge fields) localized
      to a 3-brane (``SM brane'') in the bulk (``gravity only'') 
      spacetime.
\item Bulk and boundary spacetime is flat, i.e., the bulk and
      boundary cosmological constants vanish.
\item The SM 3-brane is ``stiff''; the fluctuations of the brane 
      surface itself in the higher dimensional spacetime
      can be ignored (or, more technically, the brane fluctuations 
      have masses of order the cutoff scale)
\end{itemize}
The action for this model divides into two pieces:
\begin{equation}
S = S_{\rm bulk} + S_{\rm brane}
\end{equation}
where we are assuming for the moment that there is only one SM brane
on which all of the SM fields live.
Concentrate on the bulk action first, which is just the 
Einstein-Hilbert action for $4+n$ dimensional 
gravity:\footnote{This is my definition of $\Mstar$, and there
are plenty other others out there absorbing various factors of
$2$ and $\pi$.  One can argue that this one has the most intuitive
physical interpretation \cite{Han:2002yy}, 
and thus is the one that \emph{ought} 
to be used.  But, having not stumbled onto this until four years after 
ADD's original paper, the other definitions are unlikely to go away.}
\begin{equation}
S_{\rm bulk} = - \frac{1}{2} 
\int d^{4+n} x \sqrt{-g^{(4+n)}} \Mstar^{n+2} R^{(4+n)}
\end{equation}
We obviously integrate over all spacetime coordinates; hence
the Lagrangian has mass dimension $D = 4 + n$.  The
higher dimensional Ricci curvature scalar $R^{(4+n)}$, 
formed in the usual way from two derivatives acting on the metric,
having mass dimension 2.  This determines the mass dimension
of the coefficient of this highly relevant operator, namely
$n+2$.  

The line element of the bulk is
\begin{equation}
d s^2 = g_{M N}^{(4+n)} d x^M d x^N
\end{equation}
where we use capital letters, $M,N = 0 \ldots (4+n-1)$, as the indices 
for the bulk spacetime.  Assumption (2) of the ADD model is that spacetime 
is flat, so we can write expand $g_{M N}$ about flat spacetime 
including fluctuations.  For the moment, let's only consider 4-d metric
fluctuations, $h_{\mu\nu}$.  Then the line element is
\begin{equation}
d s^2 = \left( \eta_{\mu\nu} + h_{\mu\nu} \right) d x^\mu d x^\nu 
        - r^2 d \Omega^2_{(n)}
\end{equation}
where $d \Omega_{(n)}$ are $n$-dimensional toroidal coordinates.

Given this factorization of the metric, the higher dimensional
metric and Ricci scalar can be replaced with their four-dimensional ones,
\begin{eqnarray}
\sqrt{-g^{(4+n)}} &\rightarrow& \sqrt{-g^{(4)}} \\
R^{(4+n)} &\rightarrow& R^{(4)}
\end{eqnarray}
where $g^{(4)}$ and $R^{(4)}$ implicitly depend on $h_{\mu\nu}$.

Now rewrite the higher dimensional action in terms of 4-d modes
by ``integrating out'' the extra dimensions:
\begin{eqnarray}
S_{\rm bulk} &=& - \frac{1}{2}
                   \Mstar^{n+2} \int d^{4+n} x \sqrt{-g^{(4+n)}} R^{(4+n)} \\
             &=& - \frac{1}{2}
                   \Mstar^{n+2} \int d^4 x \int d \Omega_{(n)} r^n 
                   \sqrt{-g^{(4)}} R^{(4)} \\
             &=& - \frac{1}{2}
                   \Mstar^{n+2} (2 \pi r)^n \int d^4 x 
                   \sqrt{-g^{(4)}} R^{(4)} 
\end{eqnarray}
The last line is the action for 4-d gravity.
Matching the coefficient of the above action
with the Planck scale, one obtains the famous result
\begin{equation}
\Mpl^2 = \Mstar^{n+2} (2 \pi r)^n \; .
\label{famous-eq}
\end{equation}
This equation shows that our measured Planck scale is a derived quantity
determined by the fundamental scale of quantum gravity, $\Mstar$, 
and the volume of the extra dimensions, $V_n = (2 \pi r)^n$.

What does this result mean physically?  The weakness of 4-d long-distance
gravity is fundamentally to due the graviton being spread rather thin
across the extra dimensions with a small intersection with the SM-brane.
At long distances, gravity behaves exactly as it does in 4-d by construction, 
since we have integrated out the extra dimensions and matched this
action to the usual 4-d action.  Close to the length scale of the 
extra dimensions, however, the gravitational potential changes
and one expects to see macroscopic changes in the strength of the
force of gravity.

\subsection{Deviations from Newtonian Gravity}

Classically, the gravitational potential changes at short distance.
The Newtonian potential between two bodies of mass $m_1$ and $m_2$ is
\begin{eqnarray}
V(r') &=& \left\{ \begin{array}{lcl}
          \displaystyle{- G_N^{(4+n)} \frac{m_1 m_2}{r'^{1+n}}} 
   & \qquad & r' < r \\[1em]
          \displaystyle{- G_N \frac{m_1 m_2}{r'}} & & r' > r 
\end{array} \right.
\label{changeover-eq}
\end{eqnarray}
where $r'$ represents the distance separating the objects,
not to be confused with the size of the extra dimensions, $r$.

This leads to a vital question:  How well is gravity measured?  
The answer is, for short distances, rather poorly compared
with all the other forces.  With hindsight, we really should not
be surprised since gravity is so weak in comparison to the
other forces.  Nevertheless, it was the genius of ADD to
exploit this fact in constructing their model.

For years there have been outlandish ideas on the potential 
modification of gravity at small but macroscopic distances, 
from fifth force shenanigans to light scalar moduli from 
certain string theories.  There is some history here, and
since each particular idea has a somewhat different 
functional dependence of the strength of gravity as a 
function of distance, experimentalists simplified all this 
by parameterizing deviations in Newton's law as
\begin{equation}
V(r') = - G_N^{(4)} \frac{m_1 m_2}{r} \left( 1 + \alpha e^{-r'/\lambda} \right)
\end{equation}
The Yukawa form of the correction to Newton's law roughly
corresponds to the exchange of virtual bosons of mass
$1/\lambda$, and $\alpha=1$ corresponds to gravitational strength.

\begin{figure}[t]
\centerline{\includegraphics[width=1.0\hsize]{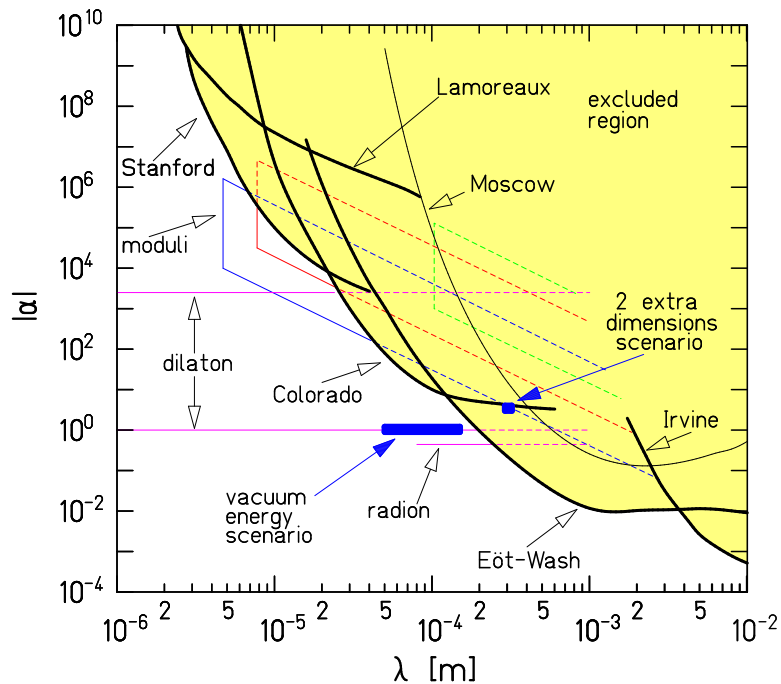}}
\caption{95\%-confidence-level constraints on ISL-violating Yukawa
interactions with 1~$\mu\mathrm{m} < \lambda < 1$ cm. The heavy curves
give experimental upper limits. 
(Fig.~5 from Ref.~\protect\cite{Adelberger:2003zx}.)}
\label{poster-fig}
\end{figure}

Fig.~\ref{poster-fig} shows how well gravity is tested at 
macroscopic distances of direct relevance to the ADD model.
You have all seen this graph many times, since it was showcased 
on the poster for this TASI school!  
This is rather busy graph showing several experimental results
as well as theoretical predictions for deviations in assorted
models.  For us, what is relevant are the experimental bounds
(i.e., ignore all of the non-capitalized identifications on the graph).  
Five experimental results are plotted that provide the strongest
constraint on deviations from Newton's law for various ranges of 
distances and strengths of forces.  For gravitational strength 
deviations, relevant to models of extra dimensions, the strongest
constraint comes from the E\"ot-Wash experiment that is consistent
with Newtonian gravity down to about 200 microns \cite{Hoyle:2004cw}.
These are remarkable experiments, and I encourage you to read
the very nice review by 
Adelberger, Heckel and Nelson \cite{Adelberger:2003zx}
for more complete details on how the experiments are done and
what they imply for the assorted models that predict deviations.

How big is the deviation in the ADD model?  The changeover
from four-dimensional gravity to higher dimensional gravity
in Eq.~(\ref{changeover-eq}) implies that once objects are brought 
to a distance $r' < r$ apart, they begin to experience 
a gravitational potential increasing in strength proportional to 
$1/r'^{n+1}$.  Gravity becomes \emph{far} stronger at short distances.
Since experiment has not found any (unambiguous) deviations 
from the $1/r'$ potential, the best we can do is to constrain
the parameters of the ADD model using these null results.

For the ADD model, it's obvious that $\alpha \sim 1$, $\lambda \sim r$
is where an $\mathcal{O}(1)$ deviation from Newtonian gravity
is expected.  Doing this a bit more carefully (for example,
see \cite{Adelberger:2003zx}), one obtains 
\begin{eqnarray}
\lambda &=& r \\
\alpha &=& \frac{4}{3} (2 n)
\end{eqnarray}
where $2 n$ sums over the number of KK gravitons with the same mass,
and the $4/3$ factor results from summing over all polarizations of 
the massive KK graviton.

Numerically, the size of this correction depends on the 
volume of the extra dimensions and the size of the fundamental 
Planck scale $\Mstar$.  
Let's take the lowest value that we could possibly imagine,
namely $\Mstar \sim 1$ TeV.  This choice ``solves'' the
hierarchy problem by lowering the cutoff scale of the Standard 
Model to 1 TeV.  There are many implications of this, 
particularly if cutoff scale effects violate the global symmetries 
of the Standard Model such as flavor, baryon number, lepton number,
etc.  This is certainly what an effective field theorist should 
\emph{expect} happens, and so ADD is immediately faced with 
serious problems.  Let's not forget, however, that every other
``solution'' to the hierarchy problem also faces the same
problems, i.e., excess violation of SM global symmetries.
(For an amusing comparison of ADD to supersymmetry, 
see \cite{Hall:2000hq}.)
Just as there are fixes to these model-induced problems in supersymmetry, 
there are also some ingenious fixes for extra dimensions, and
I'll mention a few at the end of this section and in the third lecture.

\begin{figure}[t]
\centerline{\includegraphics[width=1.0\hsize]{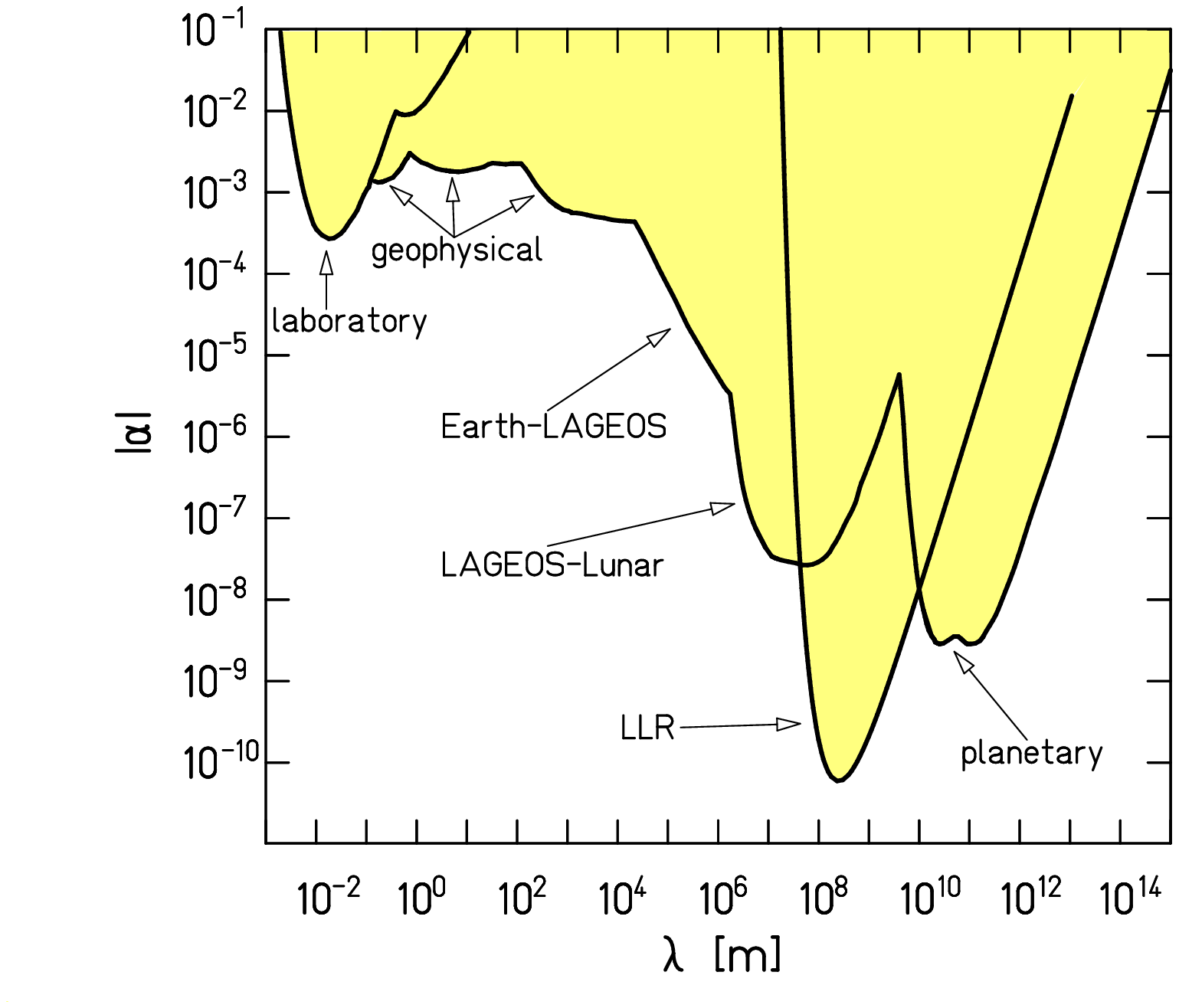}}
\caption{95\%-confidence-level constraints on ISL-violating Yukawa
interactions with $\lambda > 1$ cm. The LLR constraint is based
on the anomalous perigee precession; the remaining constraints 
are based on Keplerian tests. 
(Fig.~4 from Ref.~\protect\cite{Adelberger:2003zx}.)}
\label{planetary-fig}
\end{figure}

For now, let's just calculate.  Assuming equal-sized extra dimensions,
we can trivially solve for the radius of $n$ extra dimensions 
from Eq.~(\ref{famous-eq}),
\begin{equation}
r = \frac{1}{2 \pi} \left( \frac{\Mpl^2}{\Mstar^{n+2}} \right)^{1/n}\; .
\end{equation}
Setting $\Mstar = 1$ TeV, here is a table of the distance scale where 
one expects order one deviations from Newtonian gravity:
\begin{center}
\begin{tabular}{c|c}
number of extra dimensions & $r$ \\ \hline
$n=1$                      & $\sim 10^{12}$~m \\
$n=2$                      & $\sim 10^{-3}$~m \\
$n=3$                      & $\sim 10^{-8}$~m \\
$\vdots$                   & \\
$n=6$                      & $\sim 10^{-11}$~m \\
\end{tabular}
\end{center}
Clearly, one extra dimension $n=1$ with $\Mstar = 1$ TeV is totally
ruled out by solar system tests of Newtonian gravity.
It is amusing to see how well gravity really is measured
at these distances.  This is shown in Fig.~\ref{planetary-fig},
where estimates of the bounds on new Yukawa-like forces
are shown from a diverse set of experimental techniques 
across the distance scales in the figure.  As an aside, 
notice that for distances several orders of magnitude longer 
than the solar system, deviations from Newtonian gravity are not
well constrained.  Indeed, astronomers have in fact measured
very significant deviations from Newtonian gravity on galactic 
distance scales:  the famous mismatch between the observed rotation 
curves with what one would expect from Newtonian gravity given
the luminous mass distribution.  This is of course one motivation 
for dark matter; for reviews, see 
Refs.~\cite{Jungman:1995df,Olive:2003iq,Bertone:2004pz}.

\begin{figure}[t]
\centerline{\includegraphics[width=1.0\hsize]{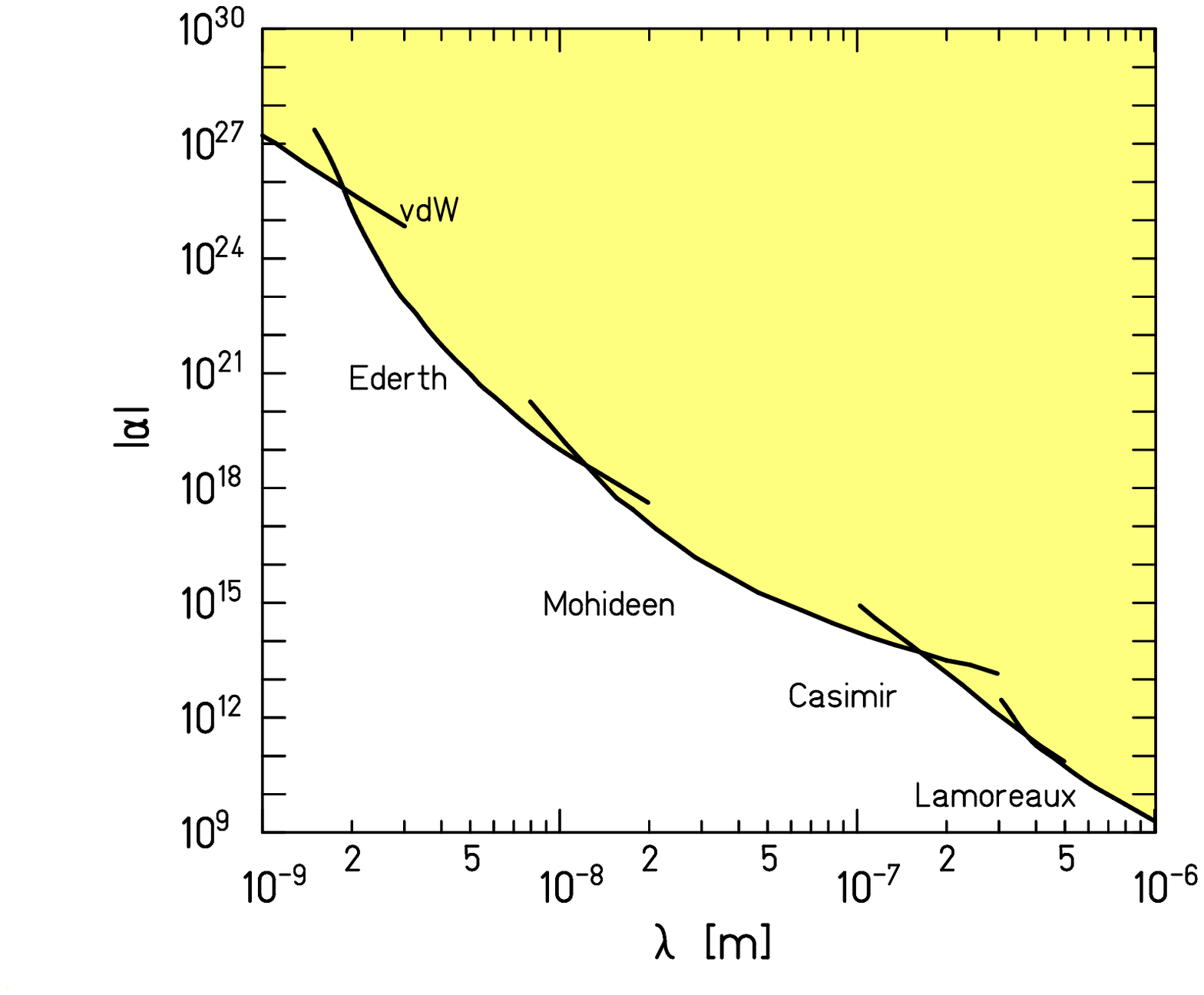}}
\caption{Constraints on ISL-violating Yukawa
interactions with $1{\rm nm}< \lambda < 1\mu$m. 
(Fig.~9 from Ref.~\protect\cite{Adelberger:2003zx}.)}
\label{nm-fig}
\end{figure}

For two extra dimensions, the predicted deviation from
Newtonian gravity occurs at $r \sim 1$ mm.  In 1998, when
ADD wrote the first paper on their model, the best experimental limit
on gravitational strength forces happened also to be at about 1 mm!
Subsequent experiments, however, have found no deviation down to
200 microns, and thus rules out two extra dimensions with a 
quantum gravity scale of $\Mstar = 1$ TeV.  For three or more
extra dimensions, the predicted deviation from Newtonian gravity
occurs at considerably smaller distances, less than ten nanometers.
The experimental constraints on new forces at these distances are
extremely weak, as shown in Fig.~\ref{nm-fig}.
The good news is that ADD with $\Mstar = 1$ TeV and $n \ge 3$ 
is not ruled by these experiments!  The bad news is that the
experiments are so far from testing gravitational strength
interactions that it is hopeless to attempt to observe
the change from $1/r$ to $1/r^{n+1}$ 
directly.\footnote{There may be ``auxiliary'' effects associated
with extra dimensions that lead to observable deviations
from Newtonian gravity, such as additional scalar moduli
that couple stronger than gravity.  Some of these effects
are shown in Fig.~\ref{poster-fig}, and I refer you to 
Ref.~\cite{Adelberger:2003zx} for details.}

\subsection{Dynamics of the Higher Dimensional Graviton}

Let's generalize the line element for arbitrary metric fluctuations
about a flat background bulk spacetime
\begin{eqnarray}
d s^2 &=& g_{MN} d x^M d x^N \\
      &=& \left( \eta_{MN} + \frac{1}{2 \Mstar^{n/2 + 1}} h_{MN} \right) 
          d x^M d x^N
\end{eqnarray}
where now $h_{MN}$ is the higher dimensional graviton.  The coefficient
is chosen for convenience to lead to canonical normalization for the
graviton.  

Insert the metric expansion into the higher dimensional 
equations of motion
\begin{equation}
G_{AB} \equiv R_{AB} - \frac{1}{2 + n} g_{AB} R = - \frac{T_{AB}}{\Mstar^{2+n}}
\end{equation}
and one obtains \cite{Giudice:1998ck}
\begin{eqnarray}
\Mstar^{n/2 + 1} G_{AB} &\equiv& 
  \Box h_{AB} 
- \partial_A \partial^C h_{CB} 
- \partial_B \partial^C h_{CA}
+ \partial_A \partial^B h_C^C \\
& &{} - \eta_{AB} \Box h_C^C
+ \eta_{AB} \partial^C \partial^D h_{CD} 
 = - \frac{T_{AB}}{\Mstar^{n/2 + 1}}
\end{eqnarray}
Define the higher dimensional coordinates as
\begin{equation}
x_M \equiv (x_\mu; y_i) \equiv (x_0, x_1, x_2, x_3; y_1, \ldots y_n)
\end{equation}
where compactification of the extra dimensions implies the $y$'s 
are periodic $y_i \ra y_i + 2 \pi r$.  Upon imposing these
periodic boundary conditions, the expansion of the higher dimensional
graviton in terms of 4-d Kaluza-Klein fields is
\begin{equation}
h_{AB}(x; y) = \sum_{m_1 =-\infty}^{\infty} \cdots \sum_{m_n =-\infty}^{\infty}
\frac{h^{(m)}_{AB}(x)}{\sqrt{V_n}} e^{i \frac{m_j y_j}{r}}
\end{equation}
where $h^{(m)}$ is a shorthand for $h^{(m_1, m_2, \ldots m_n)}$.

The SM is confined to a 3-brane within this bulk spacetime, so that
\begin{equation}
T_{AB}(x; y) = \eta^\mu_A \eta^\nu_B T_{\mu\nu} \delta^{(n)}(\vec{y})
\end{equation}
taking the SM brane to be located at $\vec{y} = 0$.
The physics content of this equation is important:
\begin{itemize}
\item We are neglecting brane fluctuations. 
\item We assume the brane is infinitely thin as represented by
the $\delta$-function of extra dimensional coordinates.  
From an effective field theory point of view the $\delta$-function 
in coordinate space is unusual, potentially leading to singularities, 
but this can be smoothed out by replacing
the $\delta$-function by $e^{-y^2 \Mstar^2}$ where the brane has a width
of at least $1/\Mstar$.  Thick-brane effects can therefore be neglected 
so long as we work in an energy regime where $\sqrt{s} \ll \Mstar$.
\end{itemize}

Now, plug in the KK expansion into Einstein's equations
\begin{eqnarray}
G_{\mu\nu}^{(k)}(x) =& f \left(h_{\mu\nu}^{(k)}, h_{\mu j}^{(k)}, h^{(k)}_{jk}
                        \right) &= - \frac{T_{\mu\nu}}{\Mpl^2} \\
G_{\mu j}^{(k)}(x) =& \ldots  &= 0 \\
G_{j k}^{(k)}(x) =& \ldots &= 0 
\end{eqnarray}
where the precise forms of $f( \, )$ are not particularly illuminating
(and can be found in \cite{Giudice:1998ck}).  Here the superscript
$(k)$ refers to the Kaluza-Klein level $k$.
Rewrite Einstein's equations
in terms of propagating (i.e., physical) degrees of freedom
for $n \not= 0$ and one obtains
\begin{eqnarray}
\left( \Box + \hat{k}^2 \right) G^{(k)}_{\mu\nu} &=& 
  \frac{1}{\Mpl} \left[ - T_{\mu\nu} 
  + \left( \frac{\partial_\mu \partial_\nu}{\hat{n}^2} + \eta_{\mu\nu} \right) 
  \frac{T_\lambda^\lambda}{3} \right] \label{KKgrav-eq} \\
\left( \Box + \hat{k}^2 \right) H^{(\vec{k})} &=& 
  \frac{1}{2 \Mpl} \sqrt{\frac{3 (n - 1)}{n + 2}} 
  T_\mu^\mu \label{KKradion-eq} \\
\left( \Box + \hat{k}^2 \right) V^{(k)}_{\mu j} &=& 0 \label{KKgravipho-eq}\\
\left( \Box + \hat{k}^2 \right) S^{(k)}_{jk} &=& 0 \label{KKscalar-eq} \; 
\end{eqnarray}
where the notation
\begin{equation}
\hat{k}^2 \equiv \sum_i^n \left| \frac{k_i}{r} \right|^2
\end{equation}
was used.
The $G^{(k)}_{\mu\nu}$ correspond to massive KK gravitons that
have absorbed one KK vector and one KK scalar for each KK level $k$.
The $V^{(k)}_{\mu j}$ correspond to massive KK graviphotons,
that absorbed one KK scalar per propagating vector per level.  
Finally the $S^{(k)}_{jk}$ and $H^{(\vec{k})}$ correspond to 
remaining massive KK scalars.  At each level (for $n > 1$), 
there is one single scalar, $H^{(\vec{k})}$, that couples to the
4-d energy momentum tensor as shown above.

These excitations are coming from the decomposition of the 
higher dimensional metric fluctuations
\begin{equation}
h_{MN} = \left( \begin{array}{c|ccc}
                h_{\mu\nu} & h_{\mu5} & h_{\mu6} & \ldots \\ \hline
                           & h_{55} & h_{56} & \ldots \\
                           &        & h_{66} & \ldots \\
                           &        &        & \end{array} \right) \; .
\end{equation}

Notice that the vectors and most scalars from this decomposition 
do not couple to SM brane-localized fields.  The gravitons and 
$H^{(\vec{k})}$ do!
Let's do an example of this in 5-d.  The five-dimensional
metric fluctuation decomposition is
\begin{equation}
\left( \begin{array}{cc}
       h_{\mu\nu} & h_{\mu5} \\
                  & \phi     \end{array} \right) 
\end{equation}
naively has KK models
\begin{equation}
\begin{array}{ccccccc}
h_{\mu\nu}^{(0)} & h_{\mu5}^{(0)} & \phantom{000} , \phantom{000} & 
h_{\mu\nu}^{(1)} & h_{\mu5}^{(1)} & \phantom{000} , \phantom{000} & \ldots \\
                 & \phi^{(0)} & \phantom{000} , \phantom{000} & 
                 & \phi^{(1)} & \phantom{000} , \phantom{000} & \ldots 
\end{array}
\end{equation}
However, a massive graviton in 4-d has five polarizations:
$h_{\mu\nu}^{(1)}$ eats $h_{\mu5}^{(1)}$ and $\phi^{(1)}$;
the latter are the longitudinal components.

It's now appropriate to go through the degree of freedom counting 
for gravitons.  As a warmup, let's begin with gauge theory.  
A general gauge field $A_M$ in $D$ dimensions has $D$ real components.  
One can always choose a gauge, such as Coulomb gauge, 
\begin{eqnarray}
\partial_M A^M &=& 0
\end{eqnarray}
reducing the number of independent components to $D-1$.
One can then do a gauge transformation on $A_M$:
\begin{eqnarray}
A_M &\ra& A_M + \partial_M \chi
\end{eqnarray}
for some real function $\chi$.  This gauge transformation
leaves the kinetic term invariant 
\begin{eqnarray}
F_{M N} &\ra& \left[   \partial_M \left( A_N + \partial_N \chi \right) 
                     - \partial_N \left( A_M + \partial_M \chi \right) 
              \right] \\
        &=& F_{M N}
\end{eqnarray}
since the $\partial_M \partial_N \chi$ terms drop out.
A massless, on-shell $D$-dimensional gauge field has therefore 
$D-2$ independent components.
A mass term for the gauge field, however, is famously known not 
to be gauge invariant since
\begin{eqnarray}
\frac{1}{2} m^2 A_M A^M &\ra& \frac{1}{2} m^2 \left[ A_M A^M
+ 2 A^M \partial_M \chi + \partial_M \chi \partial^M \chi \right] \; .
\end{eqnarray}
This gauge-transformed Lagrangian contains terms that are rather similar 
to the expansion of $D_\mu \phi D^\mu \phi$ when the minimum of $\phi$ 
is displaced from the origin.  This is just the usual Higgs mechanism 
where one can suitably reinterpret the last term as a kinetic term 
for a scalar (Goldstone) field $\chi$ and the mixing term 
represents the gauge boson/scalar mixing that gives rise to 
a gauge boson mass.  A massive, on-shell gauge field in $D$ 
dimensions therefore has $D-1$ independent components.

The story for the graviton is entirely analogous.  A $D$-dimensional
graviton is a $D \times D$ real symmetric matrix with two indices, 
and therefore $D (D + 1)/2$ components.  We can first
choose a gauge, such as the harmonic gauge, 
\begin{eqnarray}
\partial_M h^M_N &=& \frac{1}{2} \partial_N h^M_M
\label{harmonic-gauge-eq}
\end{eqnarray}
which reduces the number of independent components by $D$ since 
Eq.~(\ref{harmonic-gauge-eq}) represents $D$ independent constraints.
One can then do a general coordinate transformation on $h_{M N}$:
\begin{eqnarray}
h_{M N} &\ra& h_{M N} + \partial_M \epsilon_N + \partial_N \epsilon_M
\label{hMN-gauge-transformation-eq}
\end{eqnarray}
for some real vector function $\epsilon_N$.  This gauge transformation
leaves the kinetic term for gravity invariant.  To show this, one needs
the graviton kinetic term, namely the Einstein-Hilbert action expanded 
to leading order in the graviton fluctuation $h_{M N}$ \cite{Giudice:1998ck}
\begin{eqnarray}
R &=& - \frac{1}{2} h^{A B} \Box h_{A B} + \frac{1}{2} h^A_A \Box h^B_B
      - h^{A B} \partial_A \partial_B h^C_C 
      + h^{A B} \partial_A \partial_C h^C_B \; .
\end{eqnarray}
I leave it as an exercise to verify that $R$ is invariant under the 
general coordinate transformation Eq.~(\ref{hMN-gauge-transformation-eq}).
A massless, on-shell $D$-dimensional graviton has therefore 
$D (D - 3)/2$ independent components.  For $D=4$ we obtain
two real components, consistent with our expectations.
A mass term for the graviton field, however, is not gauge invariant
with respect to general coordinate invariance.  The Fierz-Pauli
graviton mass term in $D$ dimensions is
\begin{eqnarray}
\frac{1}{2} m^2 \left( h_{M N} h^{M N} - h_M^M h_N^N \right)
\end{eqnarray}
picks up non-zero contributions that include terms like
$h_{M N} \partial^M \epsilon^N$.  Just like for the gauge theory
example above, one can suitably reinterpret the terms of the
gauge transformed Fierz-Pauli action as analogous to a Higgs
mechanism for gravity, where now graviton-vector mixing
is the analogue of vector-scalar mixing we found above.  
Doing this analysis carefully, one finds that the vector is 
itself composed of a Goldstone (massless) vector with a Goldstone scalar, 
contributing a total $D - 1$ components to the graviton.  A massive
graviton in $D$ dimensions therefore contains $D (D - 1)/2 - 1$
components.  For example, this evaluates to 5 components for 
a 4-d graviton, matching our expectations. 
A nice description of how massive gravitons absorbs vectors,
the issues surrounding massive gravity 
(vDVZ discontinuity \cite{vDVZ}, etc.),
and what this means for a deconstruction of gravity can be
found in Ref.~\cite{deconstructinggravity}.

The potentially dangerous mode is the scalar degree of freedom
that couples to the energy momentum tensor: the radion.
The radion is a conformally-coupled scalar, and thus couples
to explicit conformal violation in the Standard Model, for example
\begin{equation}
T_\mu^\mu \sim M_W^2 W_\mu^+ W^{\mu-} + m_f \overline{f} f + \ldots \; .
\end{equation}
We have implicitly assumed that a stabilization mechanism is in 
place to fix the size of the extra dimensions and thus 
give the radion a mass sufficiently heavy so as to not modify
gravity in experimentally unacceptable ways.  This is obviously
a highly model-dependent statement, and several groups have
explored constraints on radion couplings in large extra dimension
scenarios (for example, see 
Refs.~\cite{Arkani-Hamed:1998kx,Banks:1999eg,Csaki:1999ht}).

\begin{figure}
\begin{center}
\begin{picture}(300,400)
  \Text( 20, 390 )[r]{$f(p_1)$}
  \Text( 20, 330 )[r]{$\overline{f}(p_2)$}
  \ArrowLine( 30 , 390 )( 70 , 360 )
  \ArrowLine( 70 , 360 )( 30 , 330 )
  \Photon( 70 , 361 )( 120 , 361 ){4}{5}
  \Photon( 70 , 359 )( 120 , 359 ){4}{5}
  \Text( 130 , 360 )[l]{$G_{\mu\nu}$}
  \Text( 180 , 360)[l]{$\displaystyle{-\frac{i}{4 \Mpl} 
                        \left[W_{\mu\nu} + W_{\nu\mu} \right]}$}
  \Text( 70 , 330 )[c]{(a)}
  \Text( 20, 290 )[r]{$A_\alpha(p_1)$}
  \Text( 20, 230 )[r]{$A_\beta(p_2)$}
  \Photon( 30 , 290 )( 70 , 260 ){4}{6}
  \Photon( 70 , 260 )( 30 , 230 ){4}{6}
  \Photon( 70 , 261 )( 120 , 261 ){4}{5}
  \Photon( 70 , 259 )( 120 , 259 ){4}{5}
  \Text( 130, 260 )[l]{$G_{\mu\nu}$}
  \Text( 180 ,260)[l]{$\displaystyle{-\frac{i}{\Mpl} 
     \left[ W_{\mu\nu\alpha\beta} + W_{\nu\mu\alpha\beta} \right]}$}
  \Text( 70 , 230 )[c]{(b)}
  \Text( 20, 190 )[r]{$f$}
  \Text( 20, 130 )[r]{$\overline{f}$}
  \ArrowLine( 30 , 190 )( 70 , 160 )
  \ArrowLine( 70 , 160 )( 30 , 130 )
  \Photon( 70 , 161 )( 120 , 191 ){4}{6}
  \Photon( 70 , 159 )( 120 , 189 ){4}{6}
  \Photon( 70 , 160 )( 120 , 130 ){4}{6}
  \Text( 130 , 190 )[l]{$G_{\mu\nu}$}
  \Text( 130 , 130 )[l]{$A_\alpha$}
  \Text( 180 , 160)[l]{$\displaystyle{-\frac{i}{2 \Mpl} e Q 
     \left[\gamma_\mu \eta_{\nu\alpha} + \gamma_\nu \eta_{\mu\alpha} \right]}$}
  \Text( 70 , 130 )[c]{(c)}
  \Text( 20, 90 )[r]{$g^a_\alpha(p_1)$}
  \Text( 20, 30 )[r]{$g^b_\beta(p_2)$}
  \Gluon( 30 , 90 )( 70 , 60 ){4}{6}
  \Gluon( 70 , 60 )( 30 , 30 ){4}{6}
  \Photon( 70 , 61 )( 120 , 91 ){4}{6}
  \Photon( 70 , 59 )( 120 , 89 ){4}{6}
  \Gluon( 70 , 60 )( 120 , 30 ){4}{6}
  \Text( 130 , 90 )[l]{$G_{\mu\nu}$}
  \Text( 130 , 30 )[l]{$g^c_\gamma(p_3)$}
  \Text( 180 , 60)[l]{$\displaystyle{\frac{g_3}{\Mpl} f^{abc} 
                       K(p_1, p_2, p_3)_{\mu\nu\alpha\beta\gamma}}$}
  \Text( 70 , 30 )[c]{(d)}
\end{picture}
\end{center}
\caption{Some of the Feynman rules connecting gravitons to 
SM fields, from Ref.~\cite{Giudice:1998ck}.  
Here $W^{(f)}_{\mu\nu} = (p_1 + p_2)_\mu \gamma_\nu$
and the other kinematical functions $W^{(\gamma)}_{\mu\nu\alpha\beta}$
and $K(p_1, p_2, p_3)_{\mu\nu\alpha\beta\gamma}$
can be found in \cite{Giudice:1998ck}.
Rules (b) and (c) are present for all SM groups; 
rule (d) occurs for non-Abelian groups (gluons shown).}
\label{graviton-feynman-rules-fig}
\end{figure}
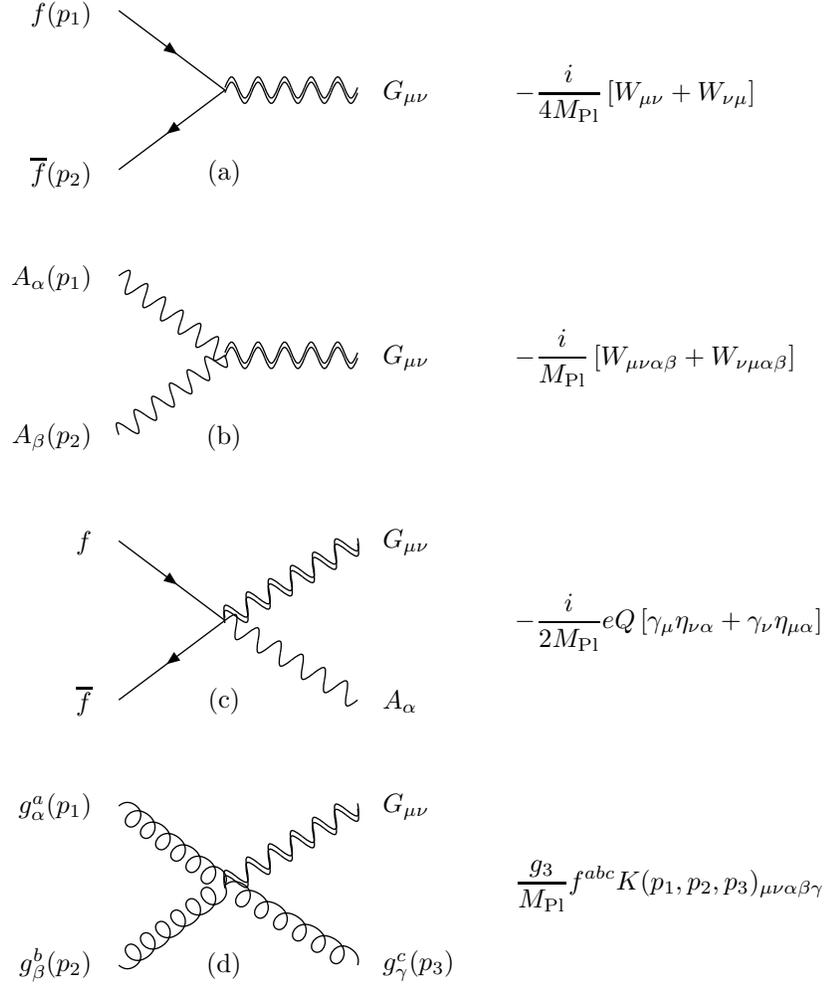

At this point, what I have done is to show you that the effects
of large extra dimensions (suitably stabilized) is reduced to 
the problem of determining the effects of the KK modes of the
graviton.  Given the results thus far, it is straightforward to
derive the Feynman rules.  I'll simply sketch the well known 
procedure for obtaining the interactions
of the KK models with matter.  The graviton couples to the 
energy momentum tensor, which we obtain from the SM action by
\begin{equation}
- 2 \frac{1}{\sqrt{-g}} 
\frac{\delta S_{\rm SM}}{\delta g_{\mu\nu}} = T^{\mu\nu}_{\rm SM}
\end{equation}
For example, the QED part of the energy-momentum tensor is
\begin{eqnarray}
T^{\mu\nu}_{\rm QED} &=& 
- F^{\mu\lambda} F_\lambda^\nu 
+ \frac{1}{4} \eta^{\mu\nu} F^{\lambda\rho} F_{\lambda\rho}
\end{eqnarray}
The Feynman rules follow directly (for example, see 
\cite{Giudice:1998ck,Han:1998sg}).  A few of them are shown in
Fig.~\ref{graviton-feynman-rules-fig}.
Given the Feynman rules, we are now ready to do phenomenology!

\subsection{Scales and Graviton Counting}

\begin{figure}[t]
\centerline{\includegraphics[width=1.0\hsize]{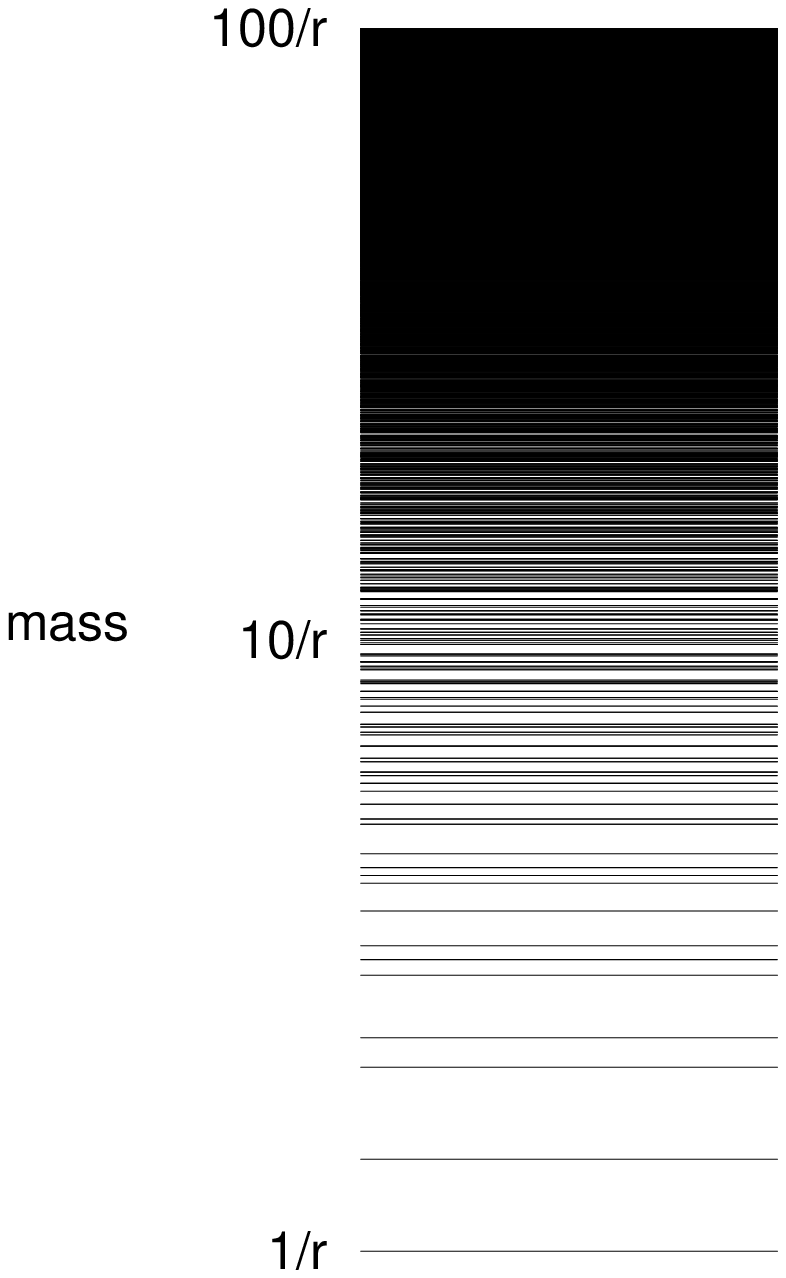}}
\caption{The mass spectrum of the KK gravitons is shown for
$n=2$.  Notice that the density of KK states fills in the energy
axis quite rapidly, allowing us to very accurately replace the
discrete set of KK states with a continuum.}
\label{dos-fig}
\end{figure}

KK gravitons have a mass $k/r$ so that the mass splittings
between KK gravitons is
\begin{equation}
\Delta m \sim \frac{1}{r} = 2 \pi \Mstar 
    \left( \frac{\Mstar}{\Mpl} \right)^{2/n} \; .
\label{r-solve-eq}
\end{equation}
Plugging in some numbers to give some feeling for the size
\begin{equation}
\Delta m \sim \left\{ \begin{array}{lcl} 
  0.003 \; \mbox{eV}  & \quad & $n=2$ \\
    0.1 \; \mbox{MeV} & & $n=4$ \\
   0.05 \; \mbox{GeV} & & $n=6$ \end{array} \right.
\qquad \mbox{for} \; \Mstar = 1 \; \mbox{TeV.}
\end{equation}
Obviously this is very small!  For illustration, 
the KK mass spectrum of gravitons for $n=2$ extra dimensions 
is shown in Fig.~\ref{dos-fig}.

It is convenient to replace the sum over modes with an 
integral over the density of states.  Fig.~\ref{dos-fig} graphically
shows that the continuum rapidly becomes an excellent approximation,
so long as experiments are not sensitive to the mass splitting.
(This is certainly true for $\Mstar$ near the TeV scale and the
number of extra dimensions is, say, $n \lsim 6$.)
The number of states $d N$ in an $n$-dimensional spatial volume
having Kaluza-Klein index between $|k|$ to $|k| + d k$ is
\begin{equation}
d N = S_{n-1} |k|^{n-1} d k \qquad S_{n-1} = \frac{2 \pi^{n/2}}{\Gamma(n/2)}
\end{equation}
where $S_{n-1}$ is the area of an $n$-sphere.  Using $m=|k|/r$,
the density of states is
\begin{equation}
d N = S_{n-1} r^n m^{n-1} d m
\end{equation}
so that the differential cross section for \emph{inclusive} 
graviton production becomes
\begin{equation}
\frac{d^2 \sigma}{d t d m} = S_{n-1} \frac{\Mpl^2}{(2 \pi)^n \Mstar^{n+2}}
m^{n-1} \frac{d \sigma_m}{d t}
\end{equation}
where $t = (p_1 - p_3)^2$; $d \sigma/d t$ is the differential
cross section for a single graviton of mass $m$; and I have
substituted for $r$ using Eq.~(\ref{r-solve-eq}).

This general formula can applied to any specific process
involving gravitons.  For example, consider real graviton production
in association with photons, $f \overline{f} \rightarrow \gamma G$.
Gravitons produced at colliders in models with large extra dimensions
escape the detector, leading to a missing energy.  Since missing energy
by itself leads to no signal at any collider detector, adding
a photon to the final state allows for ``tagging'' of large missing 
energy using the single photon signal. 
The differential cross section to produce a particular graviton is
\begin{equation}
\frac{d \sigma_m}{d t} \left( f \overline{f} \ra \gamma G \right) = 
\frac{\alpha Q_f^2}{16 N_f} \frac{1}{s \Mpl^2} 
F_1 \left( \frac{t}{s} , \frac{m^2}{s} \right)
\end{equation}
where 
$\alpha$ is the photon coupling,
$Q_f$ is the electric charge of the fermion,
$N_f$ is number of colors, and
$s$ is the center of mass energy.
Using the Feynman rules given above, the kinematical function 
$F_1(x,y)$ can be computed, and is given in the Appendix of 
Ref.~\cite{Giudice:1998ck}.
Notice that the cross section for producing a single graviton is
suppressed by the 4-d Planck scale, as you would expect.  But upon 
integrating over the huge density of states from $1/r$ up to the
energy of the process $\sqrt{s}$, one obtains 
\begin{equation}
\frac{d \sigma_m}{d t} \left( f \overline{f} \ra \gamma G \right) = 
\frac{\alpha Q_f^2}{16 N_f} \frac{1}{s \Mstar^{2+n}} 
S_{n-1} \frac{m^{n-1}}{(2 \pi)^n} 
F_1 \left( \frac{t}{s} , \frac{m^2}{s} \right)
\end{equation}
where the dependence on the 4-d Planck scale cancels!
This of course had to happen, since we could have just
as easily done the same calculation in $D$ dimensions, where
the only scale in the problem is the fundamental quantum gravity
scale, $\Mstar$, which is the coupling of the $D$-dimensional graviton!   

\begin{figure}[t]
\centerline{\includegraphics[width=1.0\hsize]{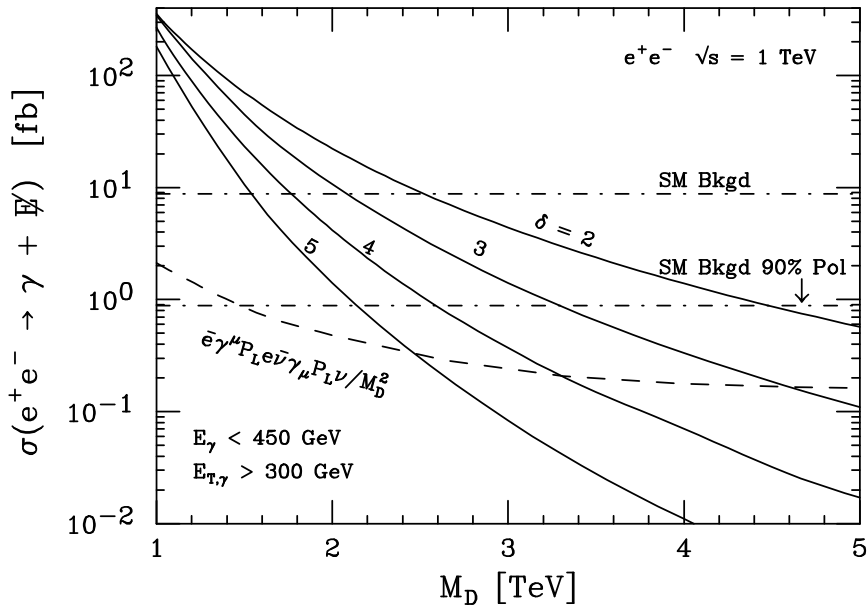}}
\caption{Total $e^+e^-\to \gamma + \mathrm{nothing}$ cross-section
at a 1 TeV center-of-mass energy $e^+e^-$ collider.  
Here $\Mstar = (2 \pi)^{-n/(2+n)} M_D \sim (0.4 \ra 0.25) M_D$ for
$n$ (called $\delta$ in the figure) between $(2 \ra 6)$.
The signal from
graviton production is presented as solid lines for various
numbers of extra dimension ($n=2,3,4,5$).  
The Standard Model background for
unpolarized beams is given by the upper dash-dotted line, and the background
with $90\%$ polarization is given by the lower dash-dotted line.  The signal
and background are computed with the requirement $E_\gamma < 450$ GeV
in order to eliminate the $\gamma Z\to \gamma\bar\nu\nu$ 
contribution to the background. The dashed line is the Standard Model 
background subtracted signal from a representative dimension-6 operator. 
(Fig.~2 from Ref.~\protect\cite{Giudice:1998ck}.)}
\label{gamma-plus-missing-fig}
\end{figure}

The total cross section is obtained by integrating over all angles.  
To give you a feeling for the size of this signal, consider the
process where the initial state particles $\overline{f}f = e^+e^-$
from a 1 TeV center-of-mass energy collider (such a linear collider
that is under active consideration by the high energy physics 
community).  The result is shown in Fig.~\ref{gamma-plus-missing-fig}.  
There are several things to glean from the figure.
Holding the energy of the incident particles fixed
(as a partonic collider does for you for free), 
you can see that more extra dimensions generically means
a smaller signal; i.e., contrast the $n=2$ curve with the $n=5$ curve.
This is easy to understand:  As the number of dimensions increases,
$1/r$ increases, and hence the density of graviton states per unit
energy interval increases dramatically.   Holding $\sqrt{s}/\Mstar$ fixed, 
then, implies that the integrated density of states between
$1/r$ to $\sqrt{s}$ always decreases as the number of dimensions
increase.  Hence, all other things considered equal, signals associated 
with graviton emission will always be harder to see as the number of 
extra dimensions increases.
Also, notice in Fig.~\ref{gamma-plus-missing-fig} that the 
estimated size of the SM background is rather substantial, and so 
one really has to get rather lucky with $\Mstar$ awfully close to
$\sqrt{s}$ to get a signal at a TeV $e^+e^-$ collider.

\begin{figure}[t]
\centerline{\includegraphics[width=1.0\hsize]{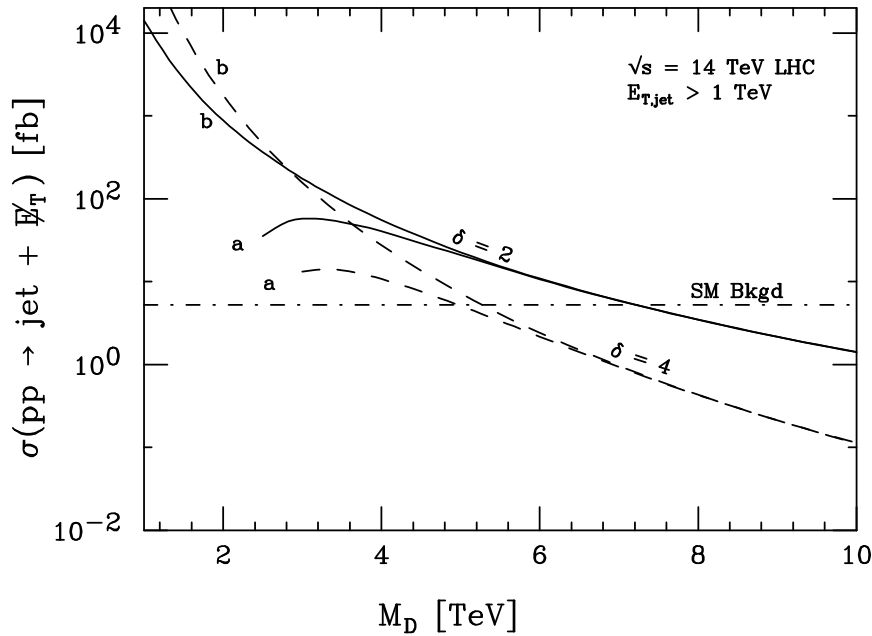}}
\caption{The total ${\rm jet}+{\rm nothing}$ cross-section
versus $M_D$ at the LHC integrated for all 
$E_{T,{\rm jet}} > 1$ TeV with the requirement that
$|\eta_{\rm jet}|< 3.0$.  
Here again $\Mstar = (2 \pi)^{-n/(2+n)} M_D \sim (0.4 \ra 0.25) M_D$ for
$n$ (called $\delta$ in the figure) between $(2 \ra 4)$.
The Standard Model background is the 
dash-dotted line, and the signal
is plotted as solid and dashed lines
for $n=2$ and $4$ extra
dimensions.  The ``a'' (``b'') lines are constructed by integrating
the cross-section over $\hat s < M_D^2$ (all $\hat s$).}
\label{hadron-ADD-fig}
\end{figure}

Hadron colliders can do much better.  This subject has 
received enormous attention (the first few papers that 
performed calculations for hadron colliders are 
\cite{Giudice:1998ck,Mirabelli:1998rt,Han:1998sg,Hewett:1998sn}).
As just one example, consider the basic partonic diagrams for the 
LHC that lead to graviton emission in association with one 
colored parton that becomes a jet in the detector.  The basic
subprocesses include $q g \ra q G$ (which gives the largest contribution),
$q \overline{q} \ra g G$, and $gg \ra g G$.  Processes with
a single quark or gluon in the final state are considered for
the same reason that a single photon in the final state was considered
for an $e^+e^-$ machine:  the large missing energy can be ``tagged''
using the monojet plus missing energy signal.  
The resulting hadronic collider process is thus
$p p \ra \mbox{jet} + \slashchar{E}_T$.
For a sufficiently high enough cut on the jet energy,
the background from one jet plus $Z \ra \overline{\nu}\nu$ 
can be sufficiently reduced to weed out the signal.
Doing the calculation in detail \cite{Giudice:1998ck}
one finds the cut on the jet energy must be in the 
several hundred GeV to TeV range.  As an illustration of the
size of this process in comparison to background, 
Fig.~\ref{hadron-ADD-fig} shows the hadronic production
cross section at leading order as a function of the 
lowered quantum gravity scale.

Another interesting signal is virtual graviton exchange.
Here one now must sum over all KK gravitons exchanged, 
so that the amplitude contains
\begin{eqnarray}
\mathcal{A} &\sim& \frac{1}{\Mpl^2} \sum \frac{1}{s - m_{\rm KK}^2} \\
            &\sim& \frac{s^{(n-2)/2}}{\Mstar^{n+2}}
\end{eqnarray}
where like before, the sum over the KK modes removes the $1/\Mpl$
suppression in favor of $1/\Mstar$ suppression.\footnote{For the case
$n=2$, the amplitude should be multiplied by $\ln s/\mu^2$.}
This result, unlike the real graviton emission in association with
photons discussed above, has certain theoretical ambiguities.

The central issue is the positive power of $\sqrt{s}$ in the numerator.  
This means that this process is ostensibly diverging as one 
approaches $\Mstar$.  This is analogous to what happens to the 
amplitude of gauge boson scattering in the Standard Model 
without a Higgs boson.  Unlike the SM, however, we don't know what
regulates quantum gravity (even if there are extra dimensions), 
and so this amplitude could well be affected by the UV physics
that smoothes out quantum gravity (strings at a TeV!).  
In effective field theory, this UV dependence corresponds to
higher dimensional operators suppressed by the cutoff scale,
i.e., the quantum gravity scale.  Hence, another way to probe 
the ADD model is to look for effects of these higher dimensional
operators.  At dimension-8, one can write the effective operator
corresponding to virtual graviton exchange at tree-level
\begin{eqnarray}
\frac{c}{\Mstar^4} \frac{1}{2} \left( T_{\mu\nu} T^{\mu\nu} - \frac{1}{n + 2}
T_\mu^\mu T_\nu^\nu \right)
\label{ADD-dim8-eq}
\end{eqnarray}
where the coefficient $c \sim 4 \pi$ is conventionally taken as
the strength of this operator at strong coupling
using naive dimensional analysis (NDA).  At dimension-6,
graviton loops can also induce new four-fermion operators, of the form
\begin{eqnarray}
c \frac{\overline{f}f\overline{f}f}{\Mstar^2} 
\label{ADD-dim6-eq}
\end{eqnarray}
where again $c \sim 4\pi$ at strong coupling.  The effects of these
operators correspond to what is usually called ``compositeness'' 
in older literature, and indeed the same analysis applies.  
Large extra dimensions are simply one realization of these operators.
The precise constraints depend on the assumption of strong coupling
(the $4\pi$ coefficient) and the particular operators in question,
but one finds numbers of order $1.5$ TeV for Eq.~(\ref{ADD-dim8-eq}) and
of order $15$ TeV for Eq.~(\ref{ADD-dim6-eq}) 
(see 2005 update of PDG \cite{Eidelman:2004wy}).

\subsection{Astrophysics}

Collider experiments can probe large extra dimensions by integrating 
over a large number of graviton modes.  To get stronger bounds
one must either increase the energy of the collider or increase
the luminosity.  The existence of light gravitons, however, 
allows a different window on this physics:  namely,
thermal systems that are hot enough to produce graviton KK modes
and large enough to produce enough of them to have an effect
on the astrophysical system.  Specifically, consider astrophysical
systems whose temperature is
\begin{equation}
T \gsim m_{\rm KK} \; .
\end{equation}
We can estimate the rate of thermal graviton production by
multiplying the coupling of each graviton by the number of 
modes accessible,
\begin{equation}
\mbox{rate of graviton production} \propto \frac{1}{\Mpl^2} 
\left( T r \right)^n \sim \frac{T^n}{\Mstar^{n+2}} \; .
\end{equation}

To find the best astrophysical bound, we want the hottest 
astrophysical system in the Universe.  The system must, however,
be well enough understood via ordinary SM physics so that we can 
use it as testing laboratory.  This system is a supernova,
and SN1987A in particular.

SN1987A is a core collapse type II supernova that went off in
our sister galaxy, the Large Magellanic Cloud, emitting a huge
amount of energy mostly into neutrinos.  The neutrinos appeared
after the stellar core of the dying star collapsed to a neutron star
and remained hot enough so that the nucleons could bremsstrahlung 
neutrinos via the weak interaction.  
Several neutrino events were recorded by
underground detectors on Earth, including Kamiokande in Japan
and IMB in the USA.  The time extent of the neutrino burst 
was several seconds, suggesting the supernova remained hot enough 
for neutrino emission to proceed on a 
macroscopic time scale.  The temperature of SN1987A is estimated 
to be $T \sim 50 \pm 20$ MeV.

SN1987A has been used to constrain all sorts of non-standard
physics (for example, see Ref.~\cite{Raffelt:1999tx}).
Here what is of most interest are the constraints on axions that 
have rather weak couplings to matter.  The basis for constraints on 
axions is that too large a coupling leads to too much axion emission
from the supernova that has the effect of providing a means to more
rapidly cool the supernova.  In this case, the time
extent of neutrino observations limits the total amount of cooling,
and thus the strength of the axion coupling.

Similarly, graviton emission can cause excess cooling of a 
supernova, and this is what we want to work out now.
The calculation is a bit different between axions and gravitons,
since axions are derivatively coupled.  The relevant reaction
for gravitons is
\begin{equation}
N + N \ra N + N + G
\end{equation}
where strong interaction effects (through pion exchange)
are unsuppressed while the nucleons $N$ themselves are 
non-relativistic.  The graviton coupling to non-relativistic
matter is
\begin{equation}
h_{\mu\nu} T^{\mu\nu} \qquad \mbox{where} \qquad 
T^{\mu\nu} = \left( \begin{array}{cc} 
                    m   & p_i \\
                    p_i & p_i p_j/m 
                    \end{array} \right)
\end{equation}
where the transverse-traceless part couples to $p_i p_j/m \sim T$,
the temperature of the non-relativistic plasma.  This causes
an extra suppression $T^2/\Mstar^2$ for the cross section to produce
gravitons compared with axions.  The thermally averaged
cross section is roughly \cite{Arkani-Hamed:1998nn}
\begin{equation}
\langle \sigma v \rangle \sim (30 \; \mathrm{mb}) 
\left( \frac{T}{\Mstar} \right)^{n+2}
\end{equation}
During SN collapse, roughly $10^{53}$ erg are released in a few seconds.
To use this to place a bound on graviton emission, we simply
require that the graviton luminosity is less than $10^{53}$ erg/s
$\sim (10^{16} \; \mathrm{GeV})^2$.  The graviton luminosity is
\begin{equation}
L_G \sim M_{\rm core} \frac{n_N^2}{\rho} (30 \; \mathrm{mb}) 
\left( \frac{T}{\Mstar} \right)^{n+2}
\end{equation} 
where $n_N$ is the nucleon number density in the core and
$\rho$ is the mass density.  
This calculation was done carefully in \cite{Cullen:1999hc},
obtaining the constraints
\begin{equation}
M_D \gsim \left\{ \begin{array}{lcl}
                  50 \; \mathrm{TeV} & \quad & n=2 \\
                   4 \; \mathrm{TeV} &       & n=3 \\
                   1 \; \mathrm{TeV} &       & n=4 
                  \end{array} \right.
\end{equation}
where $M_D = (2 \pi)^{n/(n+2)} \Mstar$.
There is no bound for $n > 4$ since the mass splitting
between the gravitons is between a few to tens of MeV,
leaving a rather small range of KK gravitons that can be emitted 
without Boltzmann suppression.

These bounds are significant for several reasons.
Perhaps the most important conclusion that can be drawn
from these results becomes manifest if we translate
the bounds on $\Mstar$ into upper bounds on the size of
the extra dimensions:
\begin{eqnarray}
r \lsim \left\{ \begin{array}{lcl}
                  10^{-4} \; \mathrm{mm} & \quad & n=2 \\
                  10^{-7} \; \mathrm{mm} &       & n=3 \\
                  10^{-8} \; \mathrm{mm} &       & n=4 
                  \end{array} \right.
\end{eqnarray}
Hence, the scale where gravitational strength deviations from
Newton's law are guaranteed to be present\footnote{To be 
distinguished from certain model-dependent effects that can
also occur in models of large extra dimensions.}
in models of large extra dimensions
is far smaller than the present experimental bound 
(about $0.2$ mm) and indeed much smaller than future experiments 
are likely able to probe (at gravitational strength). 

There is a good lesson here.  New physics can appear in myriad
experimental situations, and one must consider all of them
to obtain the best bounds on the parameters of a new physics model.
Of course this is not to suggest that continued experiments
probing gravitational deviations is futile, but
it does mean that a deviation attributed to KK gravitons 
would be (apparently) inconsistent with graviton emission from
SN1987A.  

\subsection{Cosmology}

There is another constraint that I want to discuss that
concerns excess cooling of another big astrophysical system:
the entire Universe!  
The source of cooling is the same as for supernovae, 
namely graviton emission.  In 5-d language, a 5-d graviton
can be emitted into the bulk with a coupling that is 
suppressed by just $1/\Mstar$.  In 4-d language, the probability
to emit \emph{some} KK mode goes as $1/\Mstar^{n+2}$ while
the decay of a given mode goes as $1/\Mpl$.  This means for 
high enough temperatures in the early universe, graviton emission
becomes large, and this depletes the energy of the coupled
plasma causing excess cooling that would be observed by large
differences in big bang nucleosynthesis (BBN).

The decay rate for a graviton into two photons is
\begin{equation}
\Gamma_{G \ra \gamma\gamma} = \frac{m_G^3}{80 \pi \Mpl^2}
\label{KK-grav-decay-eq}
\end{equation}
which corresponds to a decay time of
\begin{equation}
\tau \sim \left( 10^8 \; \mbox{Gyr} \right) 
  \left( \frac{\mbox{MeV}}{m_G} \right)^3
\end{equation}
Hence, once a graviton is produced it decouples from the thermal
plasma and does not decay for a long, long time.  

To extract a bound on large extra dimension models, let's compare
the ordinary Hubble expansion rate to that of cooling by gravitons.  
Cooling by Hubble expansion is roughly
\begin{equation}
\left. \frac{d \rho}{d t}\right|_\mathrm{expansion} 
\sim - 3 H \rho \sim - 3 \frac{T^2}{\Mpl^2} \rho
\end{equation}
whereas cooling by graviton emission goes as
\begin{equation}
\left. \frac{d \rho}{d t}\right|_\mathrm{evaporation} 
\sim \frac{T^n}{\Mstar^{n+2}} 
\end{equation}
These rates are equal at the ``normalcy'' temperature, which is
easily found by equating the above rates, and one obtains
\begin{equation}
T_\star = 10^{\frac{6 n - 9}{n+1}} \; \mathrm{MeV} = 
\left\{ \begin{array}{lcl}
        10 \; \mathrm{MeV} & \quad & n=2 \\
        \;\;\; \vdots & & \\
        10 \; \mathrm{GeV} &       & n=6 \; . \end{array} \right. 
\end{equation}
The normalcy temperature is is the maximum reheat temperature 
of the Universe such that cooling by ordinary Hubble expansion
dominates.  The good news is that this temperature is above
the temperature of BBN (about $1$ MeV), and so we do not expect
BBN predictions to be modified.  The bad news is that we have generally
thought that the Universe was far hotter than tens of MeV 
to tens of GeV, for example to generate weakly interacting 
dark matter, baryogenesis, inflation, etc.  All of these phenomena
need new mechanisms that operate at low temperatures.
See for example Ref.~\cite{Arkani-Hamed:1999gq} for a discussion 
of some of these issues.

\subsection{Relic Photons}

Even if the Universe is reheated to a temperature that is below 
the normalcy temperature, many light, long-lived gravitons are 
produced.  By themselves, the relic KK gravitons are not a nuisance,
but their decay products may well be.  The KK graviton decay rate
into photons was given above in Eq.~(\ref{KK-grav-decay-eq}),
and we see that some significant fraction of the KK gravitons
(ones lighter than about 5 MeV) produced in the early Universe 
will have decayed by now.  This excess source of keV to MeV photons
contributes to the diffuse cosmic $\gamma$-ray background.

\begin{figure}[t]
\centerline{\includegraphics[angle=90,width=1.0\hsize]{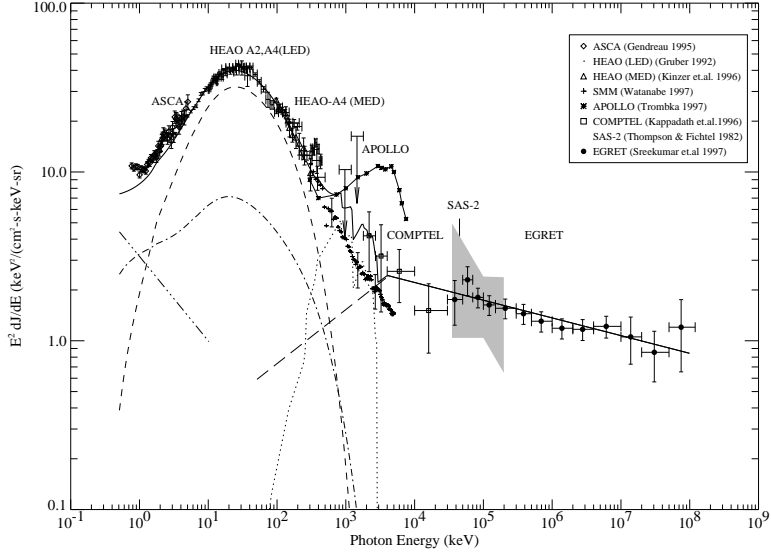}}
\caption{Multiwavelength spectrum from X-rays to $\gamma$-rays
including the revised 1$\sigma$ upper limits from the Apollo experiment
(Trombka 1997).  The thick solid line indicates 
the sum of all the components. 
(Fig.~6 from Ref.~\protect\cite{Burrows:1997eb}.)}
\label{gamma-ray-fig}
\end{figure}

Measuring the diffuse high energy photon spectrum is an ongoing 
enterprise.  The region of interest to us has been covered by
the COMPTEL experiment's $\gamma$-ray observations, shown (in conjunction 
with measurements throughout the high energy photon spectrum) in 
Fig.~\ref{gamma-ray-fig}.
Since BBN requires the Universe be reheated to at least 
about $1$ MeV, we can obtain the best bound by requiring
that the diffuse photon flux does not exceed the diffuse $\gamma$-ray
observations assuming the normalcy temperature 
$T_\star = 1$ MeV.

Ref.~\cite{Hall:1999mk} worked this out, obtaining
\begin{equation}
\left. \frac{d n_\gamma}{d E} \right|_{T_\star = 1 \; \mathrm{MeV}} = 
\alpha_n(E) \left( \frac{\mathrm{TeV}}{\Mstar} \right)^{n+2} \quad
\frac{1}{\mathrm{MeV} \; \mathrm{cm}^2 \; \mathrm{s} \; \mathrm{sr}}
\end{equation}
where the coefficient was found to be
\begin{eqnarray}
\alpha_2(4 \; \mathrm{MeV}) &\sim& 10^4 \\
\alpha_3(4 \; \mathrm{MeV}) &\sim& 0.4
\end{eqnarray}
for $n=2,3$ extra dimensions evaluated at a photon energy of $4$ MeV.
Comparing this result to the data allows us to extract a new
bound on the quantum gravity scale
\begin{equation}
M_D > \left\{ \begin{array}{rclcl}
              110 & \ra & 350 \; \mbox{TeV} & \quad & n=2 \\
                5 & \ra & 14  \; \mbox{TeV} &       & n=3 
              \end{array} \right. 
\end{equation}
where the first (second) number corresponds to restricting the
normalcy temperature to be $1$ ($2.2$) MeV.  Regardless, 
this is clearly the strongest bound we have seen on the
scale of quantum gravity from experiment for $2$ and $3$ 
extra dimensions.  No significant bound is obtained for
$n > 3$ dimensions.  

\section{Warped Extra Dimensions}

From Sundrum's lectures \cite{Sundrum:2005jf}
and Cs\'aki's lectures (with Hubisz and Meade) \cite{Csaki:2005vy}
you are already well versed on warped extra dimensions.
Their emphasis, however, is a bit different from the charge
of these lectures.  Both Sundrum and Cs\'aki were largely interested 
in warped spacetimes in which some or all of the Standard Model
fields propagate in the bulk.  This is done for all sorts of
reasons that they have expertly explained in their lectures.

In this lecture, I want to focus on the original proposal of 
Randall and Sundrum (RS) \cite{Randall:1999ee} in which only gravity 
exists in the warped extra dimension while the SM is confined to a
3-brane whose dimensionful parameters are scaled to the TeV 
scale.  This is the historical approach, which may seem a bit
dated by now, however I see this as a minimalist approach
to general topic of warped extra dimensions:  Most of what
I describe below applies to these model variants involving 
warped extra dimensions.  For instance, there are graviton resonances 
in the composite unification model \cite{compositeunification}
as well as the warped Higgsless model \cite{Csaki:2003zu}
as well as effects of radius stabilization, even if they 
are not of primary interest in those scenarios.\footnote{The 
differences between these proposals and the original 
Randall-Sundrum model is the location of the matter, Higgs, and
gauge bosons, which will however affect the strengths
of the couplings to the radion or Kaluza-Klein gravitons.}

\begin{figure}[!t]
\begin{picture}(300,140)
  \Line( 100 , 100 )( 140 , 140 )
  \Line( 100 , 20 )( 140 , 60 )
  \Line( 100 , 20 )( 100 , 100 )
  \Line( 140 , 60 )( 140 , 140 )
  \Line( 200 , 100 )( 240 , 140 )
  \Line( 200 , 20 )( 240 , 60 )
  \Line( 200 , 20 )( 200 , 100 )
  \Line( 240 , 60 )( 240 , 140 )
  \Text( 40, 78 )[c]{Planck}
  \Text( 40, 62 )[c]{Brane}
  \Text( 300, 78 )[c]{TeV}
  \Text( 300, 62 )[c]{Brane}
  \Text( 100, 10 )[c]{$y=0$}
  \Text( 200, 10 )[c]{$y=b$}
  \Text( 120, 80 )[c]{$V_{\rm Planck}$}
  \Text( 220, 80 )[c]{$V_{\rm TeV}$}
  \Text( 170, 80 )[c]{$-\Lambda$}
  \Text( 170, 130 )[c]{bulk}
\end{picture}
\caption{Sketch of the warped extra dimension RS model.}
\label{RS-setup-fig}
\end{figure}
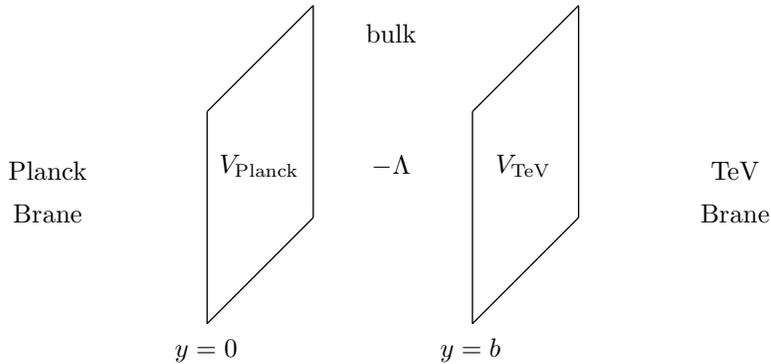

Without further ado, let's plunge into the original RS model.  
The RS model is a 5-d theory compactified on an $S^1/Z_2$ orbifold,
with bulk and boundary cosmological constants that precisely balance
to give a stable 4-d low energy effective theory with vanishing
4-d cosmological constant.  
The basic setup is sketched in Fig.~\ref{RS-setup-fig}.
The background spacetime metric is taken to be
\begin{equation}
d s^2 = e^{-k|y|} \eta_{\mu\nu} d x^\mu d x^\nu - d y^2
\end{equation}
where the metric is not as trivial as in the ADD model.
Specifically, the $y$ dependence that enters the metric 
as $e^{-k|y|}$ is known as the ``warp factor''.
The absolute value of $y$ is taken because the extra
dimension is compactified on an orbifold that identifies
$y \leftrightarrow -y$.  We will see the physical 
significance of the warp factor shortly.

The action of the model is
\begin{eqnarray}
S &=& S_{\rm bulk} + S_{\rm Planck} + S_{\rm TeV}
\end{eqnarray}
in which
\begin{eqnarray}
S_{\rm bulk} &=& - \int d^5 x \sqrt{-g} \left( \MRS^3 R - \Lambda \right) \\
S_{\rm Planck} &=& \int d^4 x \sqrt{-g_{\rm Planck}} V_{\rm Planck} \\
S_{\rm TeV} &=& \int d^4 x \sqrt{-g_{\rm TeV}} \left( V_{\rm TeV}
+ \mbox{SM Lagrangian} \right)
\end{eqnarray}
where $g_{\rm Planck}$ and $g_{\rm TeV}$ are the \emph{induced} metrics 
on the Planck and TeV branes, respectively.
Using Einstein's equations to match the metric at $y=0,b$,
one obtains
\begin{eqnarray}
V_{\rm Planck} &=& -V_{\rm TeV} = 12 k \MRS^3 \\
\Lambda &=& - k V_{\rm Planck} 
\end{eqnarray}
in terms of the AdS curvature $k$ and the fundamental
quantum gravity scale $\MRS$.\footnote{I use $\MRS$ for the 
quantum gravity scale in RS to be distinguished from $\Mstar$ 
in the ADD model.}
This solution balances bulk curvature with boundary brane
tensions (4-d cosmological constants).  There is one troubling
fact we see already, namely the brane tension of the TeV brane
is negative.  One simple way to see the effects of this is to 
phase rotate the entire TeV brane action, $S \ra -S$, that shifts
the negative sign to be in front of the kinetic terms of brane
fields.  Negative kinetic terms (for scalars, anyway) are 
well known to signal a instability in the theory in which
the kinetic terms can grow (negative) arbitrarily large.
We will ignore this issue here, and assume that the UV completion
of the RS model stabilizes the negative tension brane
(in field theory see e.g.\ Ref.~\cite{Nunes:2005up}
and in string theory see e.g.\ Ref.~\cite{Giddings:2001yu}).

Examine the SM action:
\begin{equation}
S_{\rm SM} = \int d^4 x \sqrt{-g_{\rm TeV}} \left[ 
g_{\rm TeV}^{\mu\nu} (D_\mu H)^\dag D_\nu H - \lambda (H^\dag H - v^2)^2 
+ \ldots \right]
\end{equation}
Now insert the induced metric evaluated on the TeV brane,
$(g_{\rm TeV})_{\mu\nu} = e^{-2 k b} \eta_{\mu\nu}$ 
and one obtains
\begin{equation}
\int d^4 x \left[ \eta^{\mu\nu} (D_\mu \tilde{H})^\dag (D_\nu \tilde{H}) 
- \lambda \left(\tilde{H}^\dag \tilde{H} - (e^{-k b} v)^2 \right)^2 
+ \ldots \right]
\end{equation}
in terms of the canonically normalized fields
\begin{eqnarray}
\tilde{H} &=& e^{-k b} H \\
\tilde{A}_\mu &=& e^{-k b} A_\mu \\
\tilde{f} &=& e^{-3 k b/2} f \; .
\end{eqnarray}
The central result is that the warp factor can be rescaled away
from all of the dimensionless terms of the SM (at tree-level)
by field redefinitions.  The only dimensionful operator, the Higgs 
(mass)$^2$, gets physically rescaled.  We can define a new 
Higgs vacuum expectation value (VEV)
that absorbs the warp factor 
\begin{equation}
\tilde{v} = e^{-k b} v
\end{equation}
which we see can be exponentially smaller in the canonically
normalized basis.  

The RS model presupposes that we take all fundamental mass 
parameters to be $\mathcal{O}(\Mpl)$, including $v \sim 0.1 \Mpl$.
Then for a suitably large enough slice of AdS space
relative to the curvature size, $k b \sim 35$, we obtain
\begin{equation}
\tilde{v} \sim 0.1 e^{-k b} \Mpl \sim 0.1 \; \mathrm{TeV}
\end{equation}
This is the key result that got everyone excited about the
RS model!  The interpretation is that all dimensionful
parameters on the TeV brane are ``warped'' down to TeV
scale assuming only that the size of the AdS space is 
parametrically larger than the inverse of the AdS curvature.

What are the sizes of fundamental parameters?
The 4-d effective Planck scale can be obtained by integrating
out the extra dimension to extract the coefficient of
\begin{equation}
\int d^5 x \sqrt{-g^{(4)}} e^{-2k|y|} R^{(4)}
\end{equation}
which is
\begin{equation}
\Mpl^2 = \MRS^3 \int_{y=-b}^{y=+b} e^{-2 k |y|} d y = 
\frac{\MRS^3}{k} \left( 1 - e^{-2 k b} \right) \; .
\end{equation}
Clearly, given a tiny warp factor $e^{-k b}$, all fundamental
parameters can be of order the 4-d Planck scale,
$\MRS \sim k \sim \Mpl$.

\subsection{Metric Fluctuations}

The generic ansatz for metric fluctuations about the RS background is
\begin{equation}
d s^2 = e^{-2 k |y|} g_{\mu\nu} d x^\mu d x^\nu + A_\mu d x^\mu d y 
- b^2 d y^2 \; .
\end{equation}
RS is by definition an AdS spacetime on an $S^1/Z_2$ orbifold, 
and thus the zero mode of the graviphoton, $A_\mu^{(0)}$, 
is absent.  Furthermore, we saw from ADD that in 5-d there are only
graviton KK excitations since the would-be graviphoton and graviscalar
KK excitations are completely absorbed into the longitudinal components 
of the massive KK gravitons.  The mass spectrum of RS is thus exceedingly
simple:  the massless graviton $h_{\mu\nu}^{(0)}$, the massive
KK graviton excitations $h_{\mu\nu}^{(n)}$ and a single real
scalar field $\phi$, called the radion.

Consider first just the tensor excitations; for this section, I will
follow closely Refs.~\cite{Randall:1999vf,Csaki:2004ay}.
This is easiest to 
consider when the RS metric in written in the the conformal frame
\begin{equation}
d s^2 = e^{-A(z)} \left( \eta_{\mu\nu} + h_{\mu\nu}(x,z) d x^\mu d x^\nu
 - d z^2 \right)
\end{equation}
with
\begin{equation}
e^{-A(z)} = \frac{1}{(1 + k|z|)^2} \quad , \quad A(z) = 2 \log(k|z| + 1)
\label{Az-eq}
\end{equation}
where the relationship between the two coordinates is simply
\begin{equation}
\frac{1}{1 + k|z|} = e^{-k|y|} \; .
\end{equation}


We seek linearized fluctuations about the background that satisfy
\begin{equation}
\delta G_{M N} = \frac{1}{\MRS^3} \delta T_{M N} \; .
\end{equation}
First let's fix the gauge, the ``RS gauge'', in which
the fluctuations satisfy $h_\mu^\mu = \partial_\mu h^\mu_\nu = 0$.
Expanding out $\delta G_{M N}$, keeping the leading order terms for
$h_{\mu\nu}$, one obtains 
\begin{eqnarray}
- \frac{1}{2} \partial^R \partial_R h_{\mu\nu} 
+ \frac{3}{4} \partial^R A \partial_R h_{\mu\nu} &=& 0 \; .
\end{eqnarray}
This can be written in a somewhat cleaner way by rescaling the
graviton perturbation by $h_{\mu\nu} = e^{3 A/4} \tilde{h}_{\mu\nu}$
and then the linearized Einstein equations become
\begin{equation}
- \frac{1}{2} \partial^R \partial_R \tilde{h}_{\mu\nu} + 
\left[ \frac{9}{32} \partial^R A \partial_R A
- \frac{3}{8} \partial^R \partial_R A \right] \tilde{h}_{\mu\nu} = 0 \; .
\label{RS-grav-eq-of-motion}
\end{equation}
Now separate variables
\begin{equation}
\tilde{h}_{\mu\nu}(x,z) = \hat{h}_{\mu\nu}(x) \phi(z) 
\end{equation}
where $\hat{h}$ is a four-dimensional mass eigenstate satisfying
\begin{equation}
\Box \hat{h}_{\mu\nu} = m^2 \hat{h}_{\mu\nu} \; . 
\end{equation}
The $z$-dependence within $\phi(z)$ satisfies the 1-D Schr\"odinger-like 
equation
\begin{equation}
- \partial_z^2 \phi + V(z) \phi = m^2 \phi
\label{grav-phi-eq}
\end{equation}
where the potential can be read off from Eq.~(\ref{RS-grav-eq-of-motion})
to be
\begin{equation}
V(z) = \frac{9}{16} {A'}^2 - \frac{3}{4} A''
\end{equation}
and primes denote derivatives with respect to $z$.  
Substituting for $A(z)$ using Eq.~(\ref{Az-eq}), one obtains
\begin{eqnarray}
V(z) = \frac{15}{4} \frac{k^2}{(1 + k|z|)^2} - \frac{3 k}{1 + k|z|} \delta(z)
\end{eqnarray}
where $V(z)$ is the famous ``Volcano potential'' that falls off
as $1/z^2$ far from the UV brane with a single $\delta$-function
at the caldera implying there is a single bound state 
(the massless 4-d graviton).

One can solve for the zero mode wavefunction
\begin{equation}
\phi^{(0)}(z) = e^{-3/4 A(z)} \quad \Longleftrightarrow 
\quad \phi(y) = e^{-3/4 k|y|}
\end{equation}
and thus we find that the overlap of the graviton with
the TeV brane is exponentially suppressed.  This is why gravity
appears so weak to us in the RS model.  Notice that there were no
small parameters needed to obtain this result, beyond the requirement
of the size of the extra dimension being a factor of $35$ times the
inverse of the AdS curvature.

To obtain the wavefunctions and masses of the KK modes, we first
must impose boundary conditions on the 5-d graviton wavefunction:
\begin{eqnarray}
\partial_z \phi &=& \left. - \frac{3}{2} k \phi \right|_{z=z_{\rm UV}} 
  \label{bc1-eq} \\
\partial_z \phi &=& \left. - \frac{3}{2} \frac{k}{k|z| + 1} \phi 
\right|_{z=z_{\rm IR}} \label{bc2-eq}
\end{eqnarray}
where $z_{\rm UV} = 0$ and $z_{\rm IR} = e^{k b}/k$ are the locations
of the Planck and TeV branes in the conformal coordinates.
Inserting the expression for $A(z)$ into Eq.~(\ref{grav-phi-eq}), we obtain
\begin{equation}
- \partial_z^2 \phi + \frac{15}{4} \frac{k^2}{(k|z|+1)^2} = m^2 \phi \; .
\end{equation}
The solution to this PDE can be cast in terms of a sum over
Bessel functions
\begin{equation}
\phi(z) = (k z + 1)^{1/2} \left[   a_m Y_2[m (z + 1/k)] 
                                 + b_m J_2[m (z + 1/k)] \right]
\end{equation}
where the boundary conditions Eqs.~(\ref{bc1-eq}),(\ref{bc2-eq})
completely determine the wavefunction coefficients $a_m,b_m$.
The masses of the Kaluza-Klein graviton modes are easily found
\begin{equation}
m_j = x_j k e^{-k b}
\label{mass-RS-grav-eq}
\end{equation}
where $x_j$ are the roots of the Bessel function $J_1(x_j) = 0$.
Roughly,
\begin{eqnarray}
x_j \simeq \left\{ \begin{array}{rl}
                    3.8 & \qquad j=1 \\
                    7.0 & \qquad j=2 \\
                   10.2 & \qquad j=3 \\
                   16.5 & \qquad j=4 \\
                        & \vdots 
                   \end{array} \right.
\end{eqnarray}
where the modes are obviously \emph{not} evenly spaced.
Furthermore, the quantity $k e^{-k b}$ is roughly the TeV scale
for a TeV brane, and thus these gravitons have masses of order
the TeV scale!  This is a central prediction of the RS model:
distinct spin-2 resonances with a Kaluza-Klein spectrum that
is spaced according to the roots of the first Bessel function.

What is the strength of the coupling of these gravitons?
This is where there is a big distinction between the zero mode
and the Kaluza-Klein modes of the graviton.  The zero mode
graviton has a $1/\Mpl$ strength coupling, since its wavefunction
is peaked near the Planck brane, while the Kaluza-Klein modes 
have 1/TeV strength couplings since their wavefunctions are 
peaked near the TeV brane.  This is easy to see by simply 
evaluating
\begin{equation}
\frac{\phi(z)|_{z=z_{\rm IR}}}{\phi(z)|_{z=z_{\rm UV}}} \sim e^{k b}
\end{equation}
that is exponentially enhanced by the (inverse) warp factor.
Hence, the full action for the graviton fluctuations interacting
with matter on the TeV brane is
\begin{equation}
\mathcal{L}_{\rm TeV} = - \frac{1}{\Mpl} T^{\mu\nu} h_{\mu\nu}^{(0)}
- \frac{1}{\Mpl e^{-k b}} T^{\mu\nu} \sum_{n=1}^{\infty} h_{\mu\nu}^{(n)} \; .
\end{equation}

\subsection{Phenomenology of Kaluza-Klein Gravitons}

We found that the Kaluza-Klein excitations of the graviton 
have $\mathcal{O}({\rm TeV})$ masses with $\mathcal{O}(1/{\rm TeV})$
strength couplings to fields on the TeV brane.  
This means that they will behave like new
spin-2 resonances that can be produced and observed individually.
In particular, Kaluza-Klein gravitons will decay on a timescale of
order 1/TeV.  This we can easily estimate on dimensional grounds to be
\begin{equation}
\Gamma_G \sim n_{\rm SM} \frac{m_g^3}{(\Mpl e^{-k b})^2}
\end{equation}
where $n_{\rm SM}$ is the number of SM particles into which
the graviton could decay.  

\begin{figure}[t]
\centerline{\includegraphics[angle=90,width=0.95\hsize]{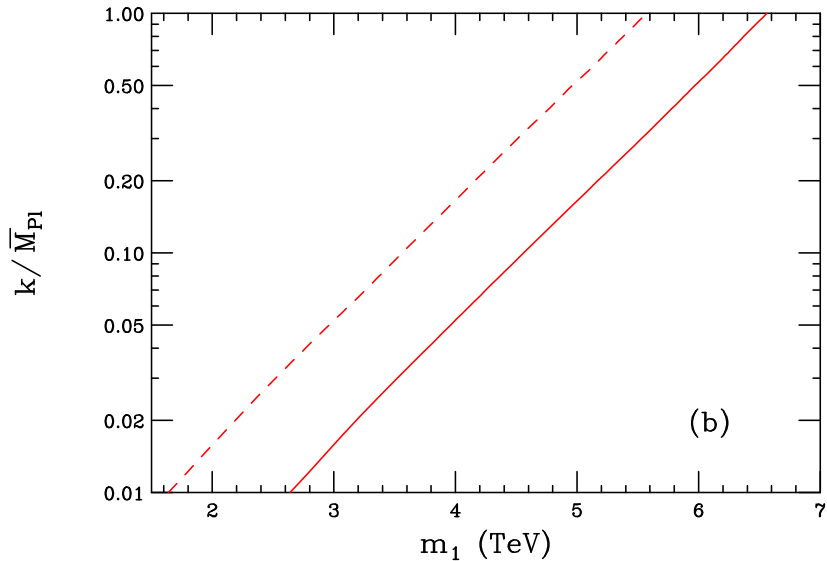}}
\caption{Exclusion regions for resonance production of the first KK graviton
excitation at the the LHC.  The dashed, solid curves correspond to
10, 100 fb$^{-1}$. The excluded region lies above and to the left 
of the curves.
(Fig.~1(b) from Ref.~\protect\cite{Davoudiasl:1999jd}.)}
\label{RS-LHC-fig}
\end{figure}

There are two parameters that determine the graviton production
cross sections and decay rates:  the warp factor and the AdS 
curvature.  The warp factor can be equivalently replaced by the 
mass of the first graviton resonance.
At the LHC, the $s$-channel exchange of gravitons leads to
new resonances, analogous to the search for new $Z'$ gauge 
bosons.  The subprocesses for this include 
$q\overline{q},gg \ra G^{(1)} \ra \ell^+\ell^-$ which provides
relatively clean leptonic events as well as
$q\overline{q},gg \ra G^{(1)} \ra q\overline{q},gg$ which leads
to pairs of jets that will reconstruct to the graviton mass.
Using these parton processes to calculate the rate of graviton
production, one can calculate a constraint on the warped
Planck scale as a function of the AdS curvature.  This is shown
in Fig.~\ref{RS-LHC-fig}, where the warped Planck scale
is related to the first graviton mass by Eq.~(\ref{mass-RS-grav-eq}).
The LHC is thus able to probe to several TeV.

\begin{figure}[t]
\centerline{\includegraphics[angle=90,width=0.95\hsize]{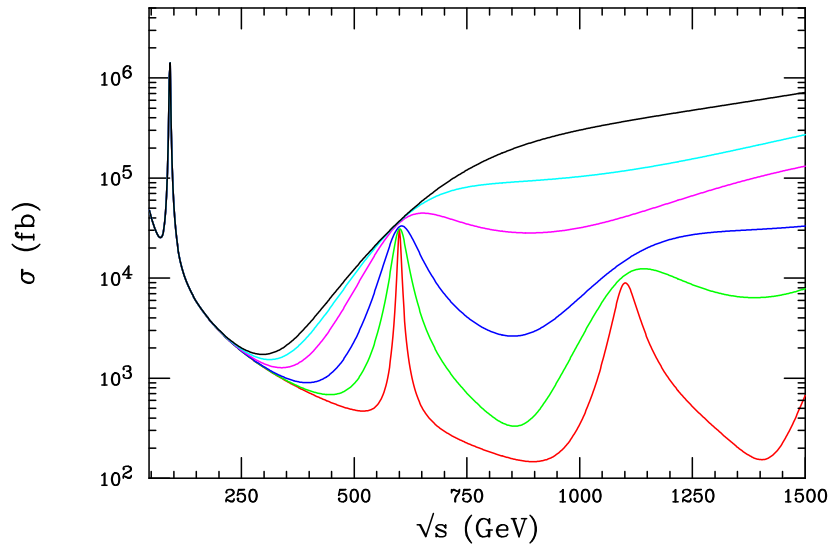}}
\caption{The cross section for $e^+e^-\to\mu^+\mu^-$ including the exchange of
a tower of KK gravitons, taking the mass of the first mode to be 600 GeV,
as a function of $\sqrt s$.  From top to bottom the curves correspond to
$k/\Mpl=1.0,\, 0.7,\, 0.5,\, 0.3,\, 0.2,\, 0.1$. 
(Fig.~2 from Ref.~\protect\cite{Davoudiasl:1999jd}.)}
\label{RS-LC-fig}
\end{figure}

At a hypothetical $e^+e^-$ collider, the broad resonances of $s$-channel 
KK graviton exchange are clearly visible (if the energy of the 
collider is high enough), allowing a detailed study of the graviton 
properties.  This is shown in Fig.~\ref{RS-LC-fig}, where an 
example of graviton production and 
decay is shown for the process $e^+e^- \ra \mu^+\mu^-$ at a
rather optimistic mass for the first KK resonance.  In any case,
the total cross section plot is rather striking!

\subsection{Radius Stabilization}

As it stands, the radion mass in the the RS model vanishes. 
This is because the brane tensions were tuned to 
balance the bulk cosmological constant.
As there is no energy cost to increase or decrease
the size of the extra dimension, radial excitations are 
uninhibited.  This is a disaster, since massless scalar fields
which couple with gravitational strength or stronger are ruled out
by light-bending and other solar system tests of general relativity.

However, there is a simple modification of the original RS proposal 
in which \emph{dynamics} can be used to stabilize the extra
dimension.  Goldberger and Wise realized that adding a bulk scalar
field with an expectation value that had a profile, namely
$v_{\rm bulk}(y)$, would be sufficient \cite{Goldberger:1999uk}.
Roughly speaking,
the bulk kinetic term wants to maximize the size of the extra dimension
whereas the bulk mass term wants to minimize the size.  
The balance between these two forces results in a stabilized 
warped extra dimension.

A complete analysis of radius stabilization is a rather intricate
affair.  The original Goldberger-Wise paper \cite{Goldberger:1999uk}
provides a very clear account
of the problem and its solution.  There are, however, a few subtleties
concerning how they implemented their solution.  Specifically,
a naive ansatz for the radion field was used which ignores both 
the radion wavefunction and the backreaction of the stabilizing 
scalar field on the metric.  This was remedied in 
Ref.~\cite{Csaki:2000zn} (see also Ref.~\cite{Tanaka:2000er}),
were we combined the metric ansatz of Ref.~\cite{Charmousis:1999rg} 
that solves Einstein's equations, with a bulk scalar potential 
that includes backreaction effects that was given by 
Ref.~\cite{DeWolfe:1999cp}.  Rather than repeating the analysis
in these lectures, let me simply quote the main results here and 
refer interested readers to the original literature for 
details on the derivation.

The main properties of the radion relevant to phenomenology
are its mass and couplings.  The radion mass is given by
\begin{eqnarray}
m_\phi &\sim& \epsilon \frac{1}{\sqrt{k b}} \Mpl e^{-k b}
\label{radion-mass-eq}
\end{eqnarray}
where $\epsilon$ is parameter that characterizes the size
of the backreaction on the metric due to the bulk scalar field
VEV profile.  In Ref.~\cite{Csaki:2000zn}, our analysis 
computed the mass of the radion treating the backreaction
as a perturbation, and thus Eq.~(\ref{radion-mass-eq}) is valid 
for $\epsilon \lsim 1$.  The key result here is that the
radion mass is of order the warped Planck scale, i.e., the 
TeV scale, but parametrically smaller by a factor $1/\sqrt{k b}$.
This is unlike the KK gravitons that we found above, where
the first KK mode had a mass of $\simeq 4 k e^{-k b}$. 
This suggests that the radion is the lightest mode in the
RS model, and thus possibly the most important excitation 
to study at colliders.  

The couplings of the radion to matter are also of vital importance.
They were first given in \cite{Csaki:1999mp,Goldberger:1999un}
and confirmed using the formalism described above in
\cite{Csaki:2000zn}.  At leading order in the radion field,
the radion couplings are
\begin{eqnarray}
\frac{\phi}{\sqrt{6} \Mpl e^{-k b}} T_\mu^\mu
\label{radion-coupling-eq}
\end{eqnarray}
where $T_\mu^\mu$ is the trace of the energy-momentum tensor
of the Standard Model.  This coupling is precisely the same
as a conformally coupled scalar field, which is not surprising
given the holographic interpretation of the RS model 
\cite{Verlinde:1999fy,Arkani-Hamed:2000ds,Rattazzi:2000hs}.
There is a fascinating story here, based on the AdS/CFT correspondence, 
concerning the duality between the 5-d RS model and a 4-d 
strongly-coupled conformal field theory.
This would take me well beyond the scope of these lectures,
and so I refer you to some excellent TASI lectures 
for details \cite{AdSCFTlectures}.
In any case, a phenomenologically successful RS model
must have dynamics that stabilize the extra dimension and
give the radion a mass, and this can be seen in either
the 5-d AdS picture or the 4-d CFT dual picture.

\subsection{Radion Phenomenology}

Given that the radion mass, Eq.~(\ref{radion-mass-eq}), is of order
(slightly smaller than) the warped Planck scale, and the couplings
of the radion are dimension-5 operators suppressed by the warped 
Planck scale, Eq.~(\ref{radion-coupling-eq}), the radion clearly
has observable effects at high energy colliders.  The mass of the
radion is effectively a free parameter $m_\phi$, which can be
bounded from above by the warped Planck scale 
\begin{eqnarray}
\Lambda = \sqrt{6} \Mpl e^{-k b}
\end{eqnarray}
and bounded from below by naturalness, namely quadratically divergent
radiative corrections to its mass.  Roughly, then, we anticipate
\begin{eqnarray}
\frac{\Lambda}{4 \pi} \lsim m_\phi \lsim \Lambda
\end{eqnarray}
which corresponds to perhaps tens of GeV up to a TeV or so
assuming $\Lambda \sim 1$ TeV.

\begin{figure}[t]
\begin{center}
\begin{picture}(300,300)
  \Text(  75, 300 )[c]{\underline{radion couplings}}
  \Text( 215, 300 )[c]{\underline{Higgs couplings}}
  \Text( 5, 280 )[r]{$f$}
  \Text( 5, 220 )[r]{$\overline{f}$}
  \ArrowLine( 10, 280 )( 40, 250 )
  \ArrowLine( 40, 250 )( 10, 220 )
  \DashLine( 40, 250 )( 80, 250 ){4}
  \Text( 85, 250 )[l]{$\phi$}
  \Text( 80, 220 )[c]{$\displaystyle{-i \frac{m_f}{\Lambda}}$}
  \Text( 155, 280 )[r]{$f$}
  \Text( 155, 220 )[r]{$\overline{f}$}
  \ArrowLine( 160, 280 )( 190, 250 )
  \ArrowLine( 190, 250 )( 160, 220 )
  \DashLine( 190, 250 )( 230, 250 ){4}
  \Text( 235, 250 )[l]{$h$}
  \Text( 240, 220 )[c]{$\displaystyle{-i \frac{m_f}{v}}$}
  \Text( 5, 180 )[r]{$W^+_\mu$}
  \Text( 5, 120 )[r]{$W^-_\nu$}
  \Photon( 10, 180 )( 40, 150 ){4}{4}
  \Photon( 40, 150 )( 10, 120 ){4}{4}
  \DashLine( 40, 150 )( 80, 150 ){4}
  \Text( 85, 150 )[l]{$\phi$}
  \Text( 80, 120 )[c]{$\displaystyle{-i \frac{2 M_W^2}{\Lambda} 
                        \eta_{\mu\nu}}$}
  \Text( 155, 180 )[r]{$W^+_\mu$}
  \Text( 155, 120 )[r]{$W^-_\nu$}
  \Photon( 160, 180 )( 190, 150 ){4}{4}
  \Photon( 190, 150 )( 160, 120 ){4}{4}
  \DashLine( 190, 150 )( 230, 150 ){4}
  \Text( 235, 150 )[l]{$h$}
  \Text( 240, 120 )[c]{$\displaystyle{-i \frac{2 M_W^2}{v} 
                        \eta_{\mu\nu}}$}
  \Text( 5, 80 )[r]{$g_\mu$}
  \Text( 5, 20 )[r]{$g_\nu$}
  \Gluon( 10, 80 )( 40, 50 ){4}{4}
  \Gluon( 40, 50 )( 10, 20 ){4}{4}
  \DashLine( 40, 50 )( 80, 50 ){4}
  \Text( 85, 50 )[l]{$\phi$}
  \Text( 80, 20 )[c]{$\displaystyle{-i \frac{b_3 \alpha_3}{2 \pi \Lambda} 
              \left( p_1 p_2 \eta_{\mu\nu} - {p_1}_\mu {p_2}_\nu \right)}$}
  \Text( 155, 80 )[r]{$g_\mu$}
  \Text( 155, 20 )[r]{$g_\nu$}
  \Gluon( 160, 80 )( 180, 60 ){4}{3}
  \Gluon( 180, 40 )( 160, 20 ){4}{3}
  \ArrowLine( 180, 60 )( 180, 40 )
  \ArrowLine( 180, 40 )( 200, 50 )
  \ArrowLine( 200, 50 )( 180, 60 )
  \DashLine( 200, 50 )( 230, 50 ){4}
  \Text( 195, 63 )[c]{$t$}
  \Text( 235, 50 )[l]{$h$}
  \Text( 240, 20 )[c]{$\displaystyle{-i \frac{(\sim \, 1) \alpha_3}{2 \pi v} 
              \left( p_1 p_2 \eta_{\mu\nu} - {p_1}_\mu {p_2}_\nu \right)}$}
\end{picture}
\end{center}
\caption{The leading order radion couplings to several fields in
the Standard Model are shown.  At tree-level, radion couplings are 
identical to Higgs boson couplings up to an overall factor $v/\Lambda$.
At loop-level, the conformal couplings of the radion are manifest,
for example, in the proportionality to the beta function coefficients
illustrated by the coupling to gluons.}
\label{radion-couplings-fig}
\end{figure}
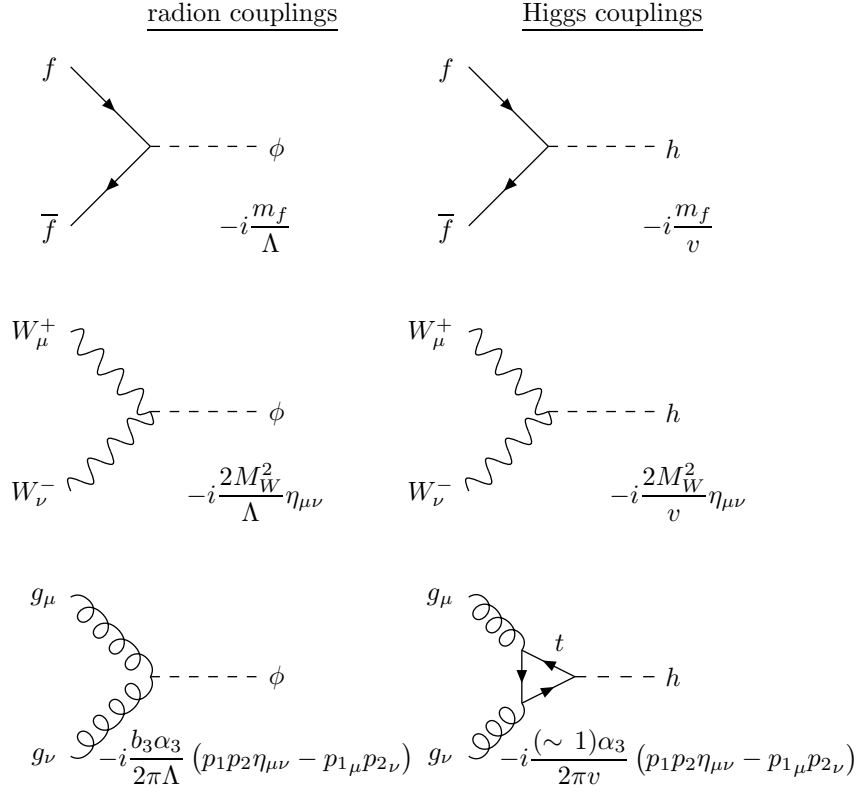

The conformal couplings of the radion are straightforward to work out.
The Standard Model in an unbroken electroweak vacuum, i.e., without a
Higgs boson, is classically scale invariant and so $T_\mu^\mu$ 
vanishes classically.  However, once a Higgs boson is introduced
and electroweak symmetry is broken, the radion couples at tree-level
to the conformal violation, namely, everywhere the Higgs VEV enters
(as well as the Higgs (mass)$^2$ term).  Scale invariance of the SM
is broken at the quantum level, since for example coupling constants 
change with energy scale under the renormalization group.
The radion also couples to this (loop-level) breaking of scale invariance.
Thus, the tree-level couplings of the radion are identical to the 
Higgs boson of the SM, except that the coupling is universally scaled 
by a factor $v/\Lambda$.  At loop-level, the scaling factor $v/\Lambda$ 
remains, but the coefficients change since the Higgs boson is not a 
conformally coupled scalar!  The radion couplings, as well as 
the couplings of the SM Higgs boson for comparison, are shown in
Fig.~\ref{radion-couplings-fig}.

There are several comments to make about these couplings.
First, the coefficient $v/\Lambda$ is expected to be of order
$1/10$ for $\Lambda \sim \; {\rm TeV}$ scale.  This drops out
of branching ratios, and thus at tree-level the radion's
branching ratios are the same as (tree-level) Higgs branching ratios.  
This suggests it is rather hard to tell the radion and the Higgs apart!  
The definitive measure is total width:  again, at tree-level 
we would expect $\Gamma_\phi = \frac{v^2}{\Lambda^2} \Gamma_h$.
Unfortunately, measuring the total width of the Higgs boson or radion
directly is somewhere between hard to really hard.
How hard it is depends on the mass of the scalar in question, 
and thus its width, 
and what instruments are at our disposal.  Obviously $s$-channel
production at a lepton collider would be ideal, measuring the
width in the same way that LEP measured the $Z$ width.
The $Z$, however, has an $\mathcal{O}(1)$ coupling with
the colliding leptons, whereas the Higgs and radion couple
(at tree-level) with only Yukawa strength.  This means an
$e^+e^-$ collider is hopeless:  the cross section is simply too small.
A muon collider is much more promising, if such a machine could 
actually be made to work.
\begin{figure}[t]
\centerline{\includegraphics[width=1.0\hsize]{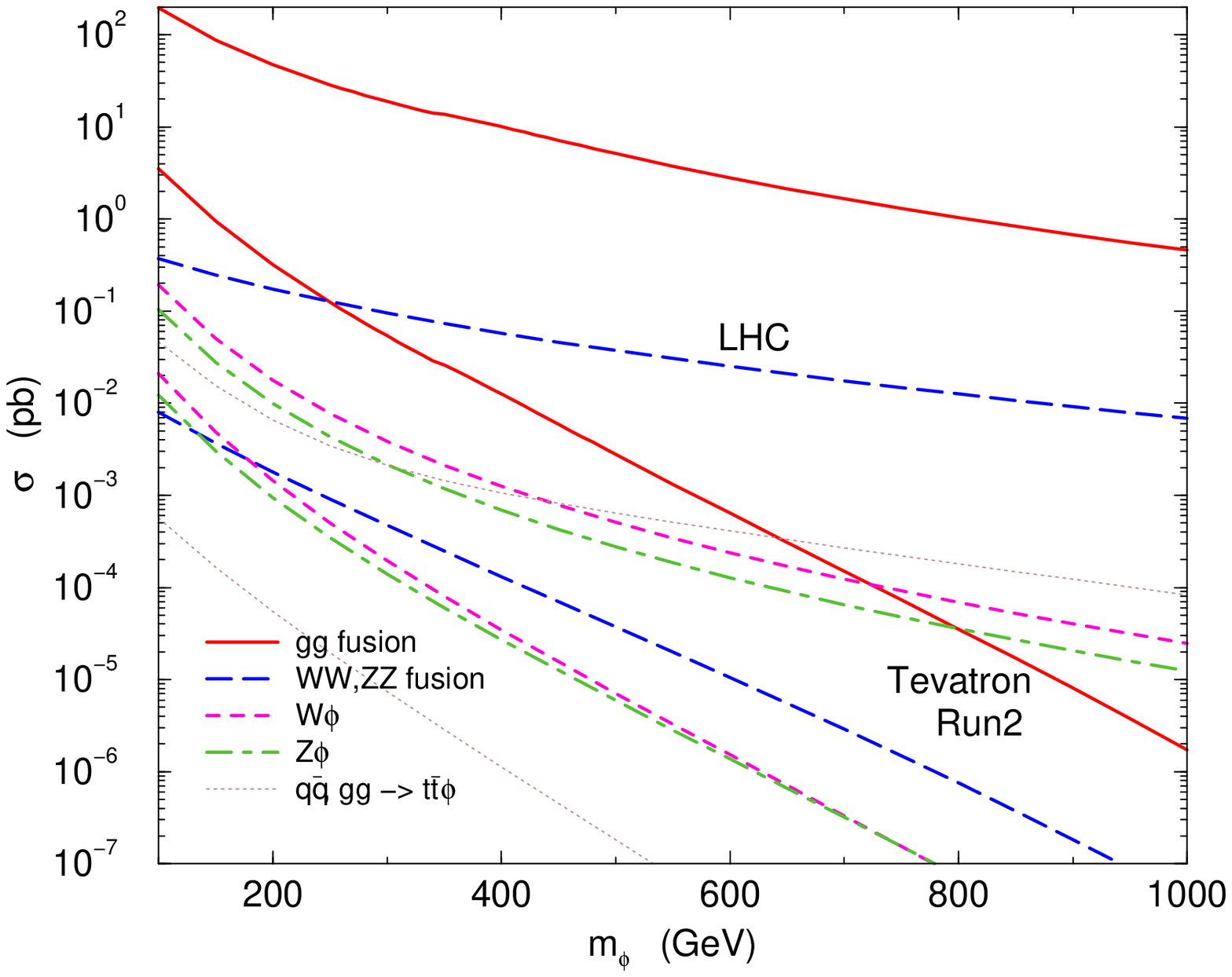}}
\caption{Production cross sections versus the mass of the radion for
$p\bar p \to \phi$ ($gg$ fusion), $p\bar p \to q q' \phi$ ($WW,ZZ$ fusion),
$p\bar p \to W\phi$, $p\bar p \to Z\phi$, and $p\bar p \to t \bar t \phi$. 
(Fig.~3(b) from Ref.~\protect\cite{Cheung:2000rw}.)}
\label{radion-production-fig}
\end{figure}

At loop-level there are differences in the branching ratios, 
and fortunately the loop effects associated with the radion couplings
are significant.  In particular, the radion coupling to gluons
has a strength that far exceeds that of the Higgs, and this has 
several important consequences.  One is that the production
cross section for radions at hadron colliders is enhanced
by roughly a factor of $b_3^2 = (7)^2$, the QCD beta function
coefficient.  This nearly compensates for the $v^2/\Lambda^2$
suppression factor from the conformal coupling, and leads
to radion production through gluon fusion that is comparable
to that of Higgs production.  The production cross
section for radion production at the LHC and the Tevatron is 
shown in Fig.~\ref{radion-production-fig} as a function
of the radion mass.

\begin{figure}[t]
\centerline{\includegraphics[width=1.0\hsize]{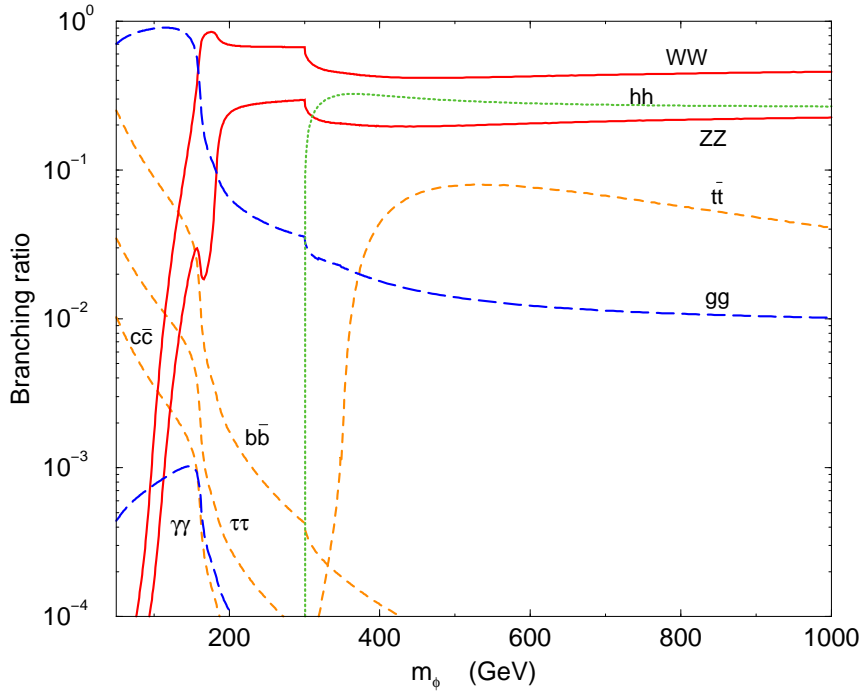}}
\caption{Branching ratios of the radion versus the radion mass. 
Here we have used $m_h=150$ GeV. 
(Fig.~2 from Ref.~\protect\cite{Cheung:2000rw}.)}
\label{radion-BR-fig}
\end{figure}

The second effect of the large coupling to gluons is that if
the radion mass is less than about $2 M_W$, the only open 
decay modes are $\phi \ra \overline{b}b$ and loop-level
decays into gluons or photons.  In the Standard Model, the
Higgs branching ratio to gluons never dominates for any range
of the Higgs boson mass.
For radions, however, the far larger coupling to gluons
dominates over the small $b$ Yukawa coupling causing the
radion to decay dominantly to gluons for the radion mass
range $20 \; {\rm GeV} \; \lsim m_\phi \lsim 2 M_W$.  This is a 
strikingly different signal as compared with a Higgs boson in 
the same mass range.
It is, in fact, a far more challenging signal to find since 
the two gluons become two jets, and the LHC is overwhelmed with
multi-jet events with modest transverse momentum.  
Branching ratios for the radion into different Standard Model
final states were calculated in \cite{Giudice:2000av,Cheung:2000rw}
and are shown in Fig.~\ref{radion-BR-fig}.

\section{Universal Extra Dimensions}

\subsection{Motivation}

We have seen that both large and warped extra dimensions
have the potential to lower the cutoff scale of the SM to
the TeV scale.  In ADD, we saw that the fundamental quantum
gravity scale is the TeV scale, and so the TeV scale becomes
the cutoff scale of the SM.  In RS, identifying the cutoff scale 
is a bit more subtle, but the potential for phenomenological
difficulties associated with higher dimensional operators 
remains (in the form of TeV brane-localized higher dimensional
operators).  This is easy to see by writing
the full effective theory for the SM plus all higher dimensional
operators on the TeV brane with natural size, i.e., of order
the 4-d Planck scale.  After the extra dimension is stabilized,
the warp factor appears everywhere there is a dimensionful
scale.  This causes the Higgs (mass)$^2$ to warp down to the
TeV scale, and simultaneously causes the higher dimensional
operators proportional to $1/\Mpl^n$ to scale ``up'' to 
$1/{\rm TeV}^n$.

In both cases, the cutoff scale is set ultimately by quantum
gravity.  It is well known lore that quantum gravity generically
violates global symmetries (e.g., black holes with Hawking evaporation).
Hence, we expect that
the presence of higher dimensional operators with dimensionful 
coefficients of order $1/\Lambda^n \sim 1/({\rm TeV})^n$ to 
violate the global symmetries of the SM.  
The global symmetries of the SM protect against an awful lot of 
curious phenomena, including $B$ and $L$ violation (together or 
separately), non-GIM flavor symmetry violation, excessive CP violation, 
custodial SU(2) violation, etc.  The SM relies on having a large
energy desert between the weak scale and the cutoff scale to
solve these problems.  
Experimental limits on the absence of these phenomena correspond 
to raising the appropriate cutoff scale of these operators high enough 
so that these phenomena do not occur.  For instance, proton decay in the 
SM occurs through
\begin{equation}
\frac{Q \overline{Q} Q \overline{L}}{M^2}
\end{equation}
among other dimension-6 $B$ and $L$ violating operators.
On dimensional grounds, this operator leads to a proton lifetime
\begin{equation}
\tau \sim \frac{M^4}{m_p^5}
\end{equation}
that must be longer than about $10^{33}$ yrs based on the
super-Kamiokande bounds \cite{Shiozawa:1998si}.
Converting this lifetime into a bound on the scale $M$ 
suppressing the operator one finds $M \gsim 10^{16}$ GeV.

Examples of other global symmetries of concern are:
\begin{itemize}
\item Lepton number violation via the dim-5 operator
\begin{equation}
\frac{(H L)^2}{M}
\end{equation}
that leads to too large (Majorana) neutrino masses
unless $M \gsim 10^{13}$ GeV.
\item Flavor-changing neutral current (FCNC) operators, such as
\begin{equation}
\frac{d \overline{s} \overline{d} s}{M^2}
\end{equation}
leading to excessive $K_0 \leftrightarrow \overline{K}_0$ mixing.
\item Baryon number violating operators, such as
\begin{equation}
\frac{Qud\overline{Q}\overline{u}\overline{d}}{M^5}
\end{equation}
leading to neutron--anti-neutron oscillations \cite{Nussinov:2001rb}.
\end{itemize}
Some of these operators, such as the one leading to 
$K_0 \leftrightarrow \overline{K}_0$ mixing, cannot simply be
forbidden by exact symmetries since this mixing has been
observed experimentally and has been successfully explained by 
GIM-suppressed flavor violation in the SM.

There are numerous proposals for solving these problems within
the contexts of the ADD model and the RS model.  
A small set of examples include
physical separation of fermions \cite{Arkani-Hamed:1999dc}, 
discrete symmetries \cite{Arkani-Hamed:1998rs,Davoudiasl:2005ks},
fermions in the bulk \cite{Sundrum:2005jf}, and so on.
Unfortunately, I don't have the time, space, or energy to review 
these ideas here.  Instead, I simply want to emphasize that 
the ADD and RS models are \emph{incomplete} as originally proposed, 
and require mechanisms to explain why these processes are small.
Universal extra dimensions have the potential to solve some
of these problems as well as provide interesting ``what if''
scenarios that can be tested in experiments.

\subsection{UED:  The Model(s)}

Universal Extra Dimensions are models in which all of the SM
fields live in $4 + n$ dimensions with the $n$ extra dimensions
taken to be flat and compact.  This basic idea has a 
long history; for some of the earlier work see \cite{otherearly}.  
In this lecture I will
primarily discuss the proposal given in \cite{Appelquist:2000nn}
and then touch on several of its numerous spin-offs:  
a solution to the proton decay problem \cite{Appelquist:2001mj};
a rationale for three generations \cite{Dobrescu:2001ae};
a test-bed for a scenario with experimental signatures that
have great similarity with (versions of) supersymmetry
\cite{Cheng:2002iz,Cheng:2002ab,Battaglia:2005zf,Datta:2005zs};
and a dark matter candidate \cite{Cheng:2002iz,Servant:2002aq}.

Promoting the full SM to extra dimensions seems like a crazy
idea for several reasons.  First, the spin-1/2 representations of the 
Poincar\'e group in higher dimensions generically have more degrees 
of freedom and differing restrictions based on chirality properties 
and anomalies (see Ref.~\cite{Lykken:1996xt} for a nice review).  
This means that fermions are generically non-chiral 
(with respect to our four dimensions).  A simple example of this
is that in five dimensions, $\gamma^5$ becomes part of the 
group structure, and so no chiral projection operators can be
constructed to reduce what are intrinsically four-dimensional 
fermion representations to chiral two-dimensional representations.  
This problem is solved by ``orbifolding'', i.e., compactifying 
on surfaces with endpoints.  In five dimensions, the only
choice is $S^1/Z_2$, which identifies opposite sides of a 
circle to create a line segment with two endpoints.
In six and higher dimensions, there are many more surfaces
to compactify on; the one that is more interesting for this
discussion is $T^2/Z_2$.

The next problem is that gauge couplings are dimensionful.
Given the higher dimensional gauge field action
\begin{equation}
S = \int d^{4 + n} x \, F_{M N} F^{M N}
\end{equation}
(where $M,N$ are the higher dimensional indices running from 
$0$ to $3 + n$), one deduces that the canonical dimension
of the gauge fields is $(2+n)/2$.  This means the gauge couplings
have dimension $-n/2$.  Gauge field couplings in higher
dimensions become analogous to graviton couplings in four dimensions,
and this means these effective theories have a cutoff of order
the scale of the coupling.  Here I have been loose with ``of order'';
a more complete accounting of the relationship between the
cutoff and the compactification scale can be found in 
Ref.~\cite{Chacko:1999hg}.
I should warn you, however, that my own partially substantiated 
hunch is that counting $4 \pi$'s in higher dimensional 
calculations to estimate the cutoff scale may be even more subtle.
This is because if one matches these higher dimensional theories 
with four-dimensional product gauge theories via deconstruction, 
it appears that the $4 \pi$ counting may be better estimated using 
just four-dimensional naive dimensional analysis.  
Anyways, for ``few extra dimensional'' theories, the difference is 
a rather innocuous $\mathcal{O}(1)$ number, and so will not really 
affect the discussion below.

Lastly, as good phenomenologists we will push the compactification
scale to the lowest possible value that is not excluded by 
experiment.  Putting gauge fields, fermions, and Higgs bosons 
in extra dimensions means there is a tower of KK excitations 
for all of these fields.  Given that we have not seen excited 
massive resonances of KK photons or gluons, while colliders have
probed up to the few hundred GeV scale, we can expect that 
$1/R \gsim$ hundreds of GeV.  We'll be much more precise below.

\subsection{Three Generations}

Dobrescu and Poppitz \cite{Dobrescu:2001ae} found a very interesting result
of promoting the SM into six dimensions.  Six dimensions
happens to be the most interesting numbers of dimensions due to 
existence of chiral fermions, as occurs for even numbers
of dimensions, and also additional anomaly cancellation constraints,
in particular due to the gravitational anomaly \cite{Alvarez-Gaume:1983ig}
that exists in two, six, ten, \ldots dimensions.

The anomalies that exist in six dimensions can be classified as
``irreducible'' gauge anomalies, 
``reducible'' gauge anomalies,
or pure or mixed gravitational anomalies.
Here ``reducible'' gauge anomalies correspond to most of the
SM ones involving $U(1)_Y$ and $SU(2)_L$.  Dobrescu and Poppitz
argue that we do not need to care about anomalies associated with 
symmetries that are spontaneously broken.  One way to rationalize
this argument is that since the electroweak breaking scale and
the compactification scale are roughly the same for UED, 
any effects associated with anomalies of these spontaneously
broken gauge symmetries can be absorbed by cutoff scale effects.
Alternatively, they also point out that the reducible gauge anomalies 
can be canceled by a higher dimensional version of the 
Green-Schwarz mechanism, see Ref.~\cite{Dobrescu:2001ae} for details.

The irreducible gauge anomalies are those associated with 
$SU(3)_c$ and $U(1)_{\rm em}$.  Since colored particles and
electrically charged particles are vector-like, the only
non-trivial anomaly comes from $U(1)_{\rm em} [SU(3)_c]^3$ 
(in six dimensions, anomalies correspond to box diagrams 
connecting four gauge fields together).  

The pure and mixed gravitational anomalies arise with respect
to the six-dimensional (6-d) chirality assignments of the 6-d 
generalizations of SM fermions.  Fermions can take on either chirality, 
using
\begin{eqnarray}
\Gamma^7 \cdot f &\rightarrow& \pm f_{\pm}
\end{eqnarray}
where $\Gamma^M$ are anti-commuting $2^{D/2} \times 2^{D/2}$
[$2^{(D-1)/2} \times 2^{(D-1)/2}$] matrices for even [old] 
$D$ dimensions.
The $f_{\pm}$ are the 6-d chiral fermions (each chirality is 
a four-component spinor) analogous to
the more familiar $f_{L,R}$ 4-d chiral fermions (where each
chirality is a two-component spinor).
Requiring the irreducible gauge anomaly to cancel
combined with the pure and mixed gravitational anomalies
leads to one of four possible 6-d chiral assignments
\begin{eqnarray}
Q_+, u_-, d_-, L_+, e_-, N_- \\
Q_+, u_-, d_-, L_-, e_+, N_+ 
\end{eqnarray}
where the other two assignments simply flip $+ \leftrightarrow -$
for all fermion species.  Already, an interesting result is 
that a gauge-neutral fermion, $N$, is required to exist so
that the pure and mixed gravitational anomaly is canceled.
This requirement is curiously similar to an analogous phenomena 
that occurs with gauged flavor symmetry extensions of the 
(four-dimensional) SM \cite{Kribs:2003jn}.

Finally, there are the global gauge anomalies.  These are higher
dimensional analogues of the Witten anomaly for SU(2) \cite{Witten:1982fp}.
Global gauge anomalies potentially exist for $SU(3)$, $SU(2)$, and $G_2$
in six dimensions.  In UED, $SU(3)_c$ is vector-like and so 
automatically cancels.  $SU(2)_L$, however, requires
\begin{equation}
n(2_+) - n(2_-) = 0 \; {\rm mod} \; 6
\end{equation}
where $n(2_\pm)$ corresponds to the number of doublets with 6-d 
chirality $\pm$.  For one generation, $n(Q) = 3$, $n(L) = \pm 1$
implying $n(+) - n(-) = 2 \; {\rm or} \; 4$.  For $n_g$ generations,
this becomes
\begin{eqnarray}
n(Q) &=& 3 n_g \\
n(L) &=& \pm n_g \\
n(2_+) - n(2_-) &=& 2 n_g \; {\rm or} \; 4 n_g
\end{eqnarray}
and we see that the global gauge anomaly is canceled with 
$n_g = 3$ (${\rm mod} \; 3$) generations!

\subsection{Proton Decay}

We already remarked on the potential problems with operators
that lead to proton decay in models with a low cutoff scale.
Ref.~\cite{Appelquist:2001mj} pointed out that
part of the global symmetry associated with the extra dimensional
coordinates can be utilized to restrict the forms of higher
dimensional operators that are allowed.  In particular,
in 6-d the Poincare symmetry $SO(1,5) \ra SO(1,3) \times U(1)_{45}$
where the $U(1)_{45}$\footnote{``$45$'' is a label for one U(1), 
not to be confused with forty-five U(1)'s.} corresponds to rotations 
between the fourth and fifth (extra dimensional) coordinates.  

Chiral 6-d fermions decompose under 4-d $SU(2)_L$ chirality as
\begin{equation}
\phi_{\pm} = \phi_{\pm L} + \phi_{\pm R}
\end{equation}
defined via the projection operators 
\begin{equation}
P_{\pm L} = P_{\mp R} = \frac{1}{2} \left( 1 \mp \Sigma^{45} \right)
\end{equation}
where $\Sigma^{\alpha\beta}/2$ are the generators of the spin-1/2
representations of $SO(1,5)$.  The 4-d chiral fermions have
$U(1)_{45}$ charges given by the eigenvalues of $\Sigma^{45}/2$:
$\mp 1/2$ for $\phi_{\pm L}$ and $\pm 1/2$ for $\phi_{\pm R}$.
This implies the $U(1)_{45}$ charge assignments for the SM fields:
\begin{center}
\begin{tabular}{c|cc} \hline\hline
fermion            & $U(1)_{45}$ charge & $U(1)_B$ \\ \hline
$Q_{+L}$           &       $-1/2$       & $1/3$ \\
$u_{-R}$, $d_{-R}$ &       $-1/2$       & $1/3$ \\
$L_{+L}$           &       $-1/2$       & $0$ \\
$e_{-R}$, $N_{-R}$ &       $-1/2$       & $0$ \\ \hline\hline
\end{tabular}
\end{center}
Baryon number violation requires three quark fields, 
but obviously no combination of three quarks is invariant 
under $U(1)_{45}$.  To obtain a $\Delta B = 1$ operator, 
therefore, we need \emph{three} lepton fields to make the
operator $U(1)_{45}$-invariant.  The lowest dimensional 
operator\footnote{There are several operators at dim-16 involving 
the singlet $N$, but for conciseness I will only consider the 
lowest dimensional operator involving SM fields.} 
occurs at dimension-17
\begin{equation}
\mathcal{O}_{17} = 
\frac{\left( \overline{L}_{+L} d_{-R} \right)^3 H^\dag}{\Lambda^{11}}
\end{equation}
After integrating over the extra dimensions, the 4-d low energy 
effective theory contains the dim-9 operator
\begin{equation}
\frac{v}{R^5 \Lambda^{11}} (\overline{\nu}_L d_R) (\overline{l}_L d_R)^2
\end{equation}
leading to proton decay via the 5-body process such as
$p \ra e^- \pi^+ \pi^+ \nu\nu$.
Putting in appropriate coefficients and form factors, one
obtains \cite{Appelquist:2001mj} 
\begin{equation}
\tau_p \simeq 10^{35} \; \mbox{yr} \, 
\left( \frac{1/R}{500 \; {\rm GeV}} \right)^{12}
\left( \frac{\Lambda R}{5} \right)^{22} \; .
\end{equation}
Hence, for $1/R$ of order the weak scale with a cutoff scale
$\Lambda \gsim 5/R$, the 6-d UED model based on 
$T^2/Z_2$ is completely safe from proton decay.

One can show more generally that the sum rule
\begin{equation}
3 \Delta B \pm \Delta L = 0 \; {\rm mod} \; 8
\end{equation}
is satisfied for all of the zero-mode fields.
This forbids: proton decay with less than 6 fermions;
$\Delta B = 2$, $\Delta L = 0$ baryon-number violating interactions 
leading to neutron--anti-neutron oscillations; and
$\Delta B = 0$, $\Delta L = 2$ lepton-number violating interactions
leading to Majorana neutrino masses.  Hence, many of the most
dangerous violations of the SM global symmetries are forbidden
or sufficiently suppressed.

\subsection{UED:  The Model}

Having piqued your interest in UED by the argument for three 
generations as well as naturally allowing a low cutoff scale,
let's now delve into the UED model and its phenomenology.

The action for the SM in higher dimensions is
\begin{eqnarray}
S &=& \int d^4 x \int d^n y \, \bigg[ 
      \frac{1}{2 \overline{g}^2} F_{MN} F^{MN} \nonumber \\
& &{} + i \overline{Q} \Gamma^M D_M Q 
      + i \overline{u} \Gamma^M D_M u
      + i \overline{d} \Gamma^M D_M d \nonumber \\
& &{} + \overline{Q} \lambda_u u i \sigma_2 H^* 
      + \overline{Q} \lambda_d d H \nonumber \\
& &{} + \mathcal{L}_{\rm Higgs} + \mbox{leptons} + \ldots \bigg]
\end{eqnarray}
where gauge interactions, Yukawa interactions, and Higgs interactions
are all \emph{bulk} interactions.  These couplings are thus 
\emph{dimensionful}, since this is a higher dimensional theory.
In particular, it must be stressed that there are no $\delta(y)$
functions present.  The UED model, by definition, has no tree-level
brane-localized fields or interactions.  

Demanding that all fields and interactions are bulk interactions,
with no $\delta$-functions in extra-dimensional coordinate space, 
has one extremely important
consequence.  To see this, let's first decompose a ($4+n$)-dimensional
gauge field into its 4-d Kaluza-Klein (KK) components:
\begin{eqnarray}
A_\mu(x,y) &=& \frac{\sqrt{2}}{(2 \pi R)^{n/2}} \Bigg\{ A_\mu^{(0)}(x) 
\nonumber \\
           & &{} + \sqrt{2} \sum_{j_1, \ldots, j_n}
A_\mu^{(j_1,\ldots,j_n)}(x) \, 
\cos \left[ \frac{j_1 y_1 + \ldots + j_n y_n}{R} \right] \Bigg\}
\end{eqnarray}
We could continue this discussion in an arbitrary number
of dimensions, but for simplicity let's concentrate on
just one extra dimension.  There is no loss of generality
to the basic argument I am about to present by specializing to 5-d.
Indeed, numerous papers that have been written about UED 
have concentrated on the 5-d version, so this sets us up
nicely to discuss this body of work.  In most cases, it is
more complicated but nevertheless straightforward to extend
the 5-d discussions into 6-d to preserve the properties that
we found in the first two subsections.

Back to the significance of the absence of $\delta$-functions
in extra-dimensional coordinate space.
This is best understood with an example.  Consider two distinct 5-d
theories:  one contains bulk fermions $F$, the other contains 
boundary fermions $f$ (localized at $y=0$), while both are coupled 
to a bulk gauge field $A_M$.  
The theory with boundary fermions has an action
\begin{eqnarray}
\int d^4 x \, d y \, \overline{f} \Gamma^M D_M f \, \delta(y) 
\end{eqnarray}
that upon KK expansion becomes
\begin{eqnarray}
\int d^4 x \, d y \, \overline{f} \Gamma^\mu \left[ A_\mu^{(0)} + 
\sqrt{2} \sum_j A_\mu^{(j)} \cos \left( \frac{j y}{R} \right) \right] 
f \, \delta(y)
\end{eqnarray}
where there is an overall constant as well as an additive
set of interactions with the fifth component of the gauge field
(an additional scalar field) that I'm not bothering about here.
Integrate out the fifth dimension, assumed to be on the interval
$S^1/Z_2$,
\begin{eqnarray}
\int_0^{\pi R} d y \, \cos\left( \frac{j y}{R} \right) \delta(y) = 1
\end{eqnarray}
and one is left with the 4-d Lagrangian
\begin{eqnarray}
\int d^4 x \, \overline{f} \Gamma^\mu 
\left[ A_\mu^{(0)} + \sqrt{2} \sum_j A_\mu^{(j)} \right] f
\end{eqnarray}
where the 4-d boundary fermions couple to all of the KK modes
with the same strength.

\begin{figure}[t]
\begin{picture}(400,100)
  \ArrowLine(50,85)(50,50)
  \ArrowLine(50,50)(50,15)
  \Photon(50,50)(100,50){4}{4}
  \Text(45,80)[r]{$F^{(0)}$}
  \Text(45,20)[r]{$F^{(0)}$}
  \Text(95,60)[r]{$A_\mu^{(0)}$}
  \Text(75,10)[c]{(a)}
  \ArrowLine(150,85)(150,50)
  \ArrowLine(150,50)(150,15)
  \Photon(150,50)(200,50){4}{4}
  \Text(145,80)[r]{$F^{(k)}$}
  \Text(145,20)[r]{$F^{(0)}$}
  \Text(195,60)[r]{$A_\mu^{(k)}$}
  \Text(175,10)[c]{(b)}
  \ArrowLine(250,85)(250,50)
  \ArrowLine(250,50)(250,15)
  \Photon(250,50)(300,50){4}{4}
  \Text(245,80)[r]{$F^{(0)}$}
  \Text(245,20)[r]{$F^{(0)}$}
  \Text(295,60)[r]{$A_\mu^{(k)}$}
  \Line(240,15)(290,85)
  \Line(240,85)(290,15)
  \Text(275,10)[c]{(c)}
\end{picture}
\caption{Example of interactions that are allowed [(a) and (b)]
and are not allowed [(c)] by KK number (or KK parity) conservation.}
\label{UED-diagrams-fig}
\end{figure}
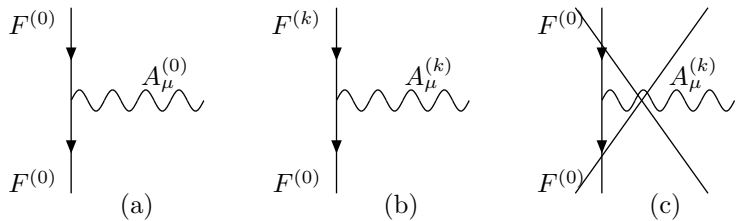

Now contrast this to what happens in the theory with only bulk fermions.  
The action
\begin{eqnarray}
\int d^4 x \, d y \, \overline{F} \, \Gamma^M D_M F 
\end{eqnarray}
is KK expanded into
\begin{eqnarray}
\int d^4 x \, d y \, \overline{F}^{(0)} \Gamma^\mu 
\left[ A_\mu^{(0)} + 
\sqrt{2} \sum_j A_\mu^{(j)} \cos \left( \frac{j y}{R} \right) \right] F^{(0)} 
\end{eqnarray}
plus all of the terms with KK excitations for the fermions that
I neglected to write here.  Integrate out the fifth dimension
\begin{eqnarray}
\frac{2}{\pi R} \int_0^{\pi R} d y \, \cos\left( \frac{j y}{R} \right) &=& 
2 \delta_{j0} 
\label{bulk-integral-1-eq}
\end{eqnarray}
where this integral vanishes for all $j$ except $j=0$.
One is left with the 4-d Lagrangian
\begin{eqnarray}
\int d^4 x \, \overline{F}^{(0)} \Gamma^\mu A_\mu^{(0)} F^{(0)}
\end{eqnarray}
where the 4-d zero mode fermions couple \emph{only} to the zero mode 
of the gauge field!  Generalizing to the $k^{\rm th}$ KK mode of one of
the bulk fermions interacting with the $j^{\rm th}$ KK mode of the gauge
field, the integral Eq.~(\ref{bulk-integral-1-eq}) becomes
\begin{eqnarray}
\frac{2}{\pi R}
\int_0^{\pi R} dy \, \cos\left( \frac{j y}{R} \right) 
                     \cos\left( \frac{k y}{R} \right) = \delta_{jk}
\end{eqnarray}
leading to the non-zero 4-d interactions
\begin{eqnarray}
\int d^4 x \, \overline{F}^{(k)} \Gamma^\mu A_\mu^{(k)} F^{(0)}
+ \overline{F}^{(0)} \Gamma^\mu A_\mu^{(k)} F^{(k)}\; , 
\end{eqnarray}
also shown diagrammatically in Fig.~\ref{UED-diagrams-fig}.
The presence of interactions with an even number of same-level KK modes
is due precisely to the absence of $\delta$-functions.  
The $\delta$-functions are sources for brane-localized interactions 
which completely break translation invariance in the fifth dimension.  
The absence of $\delta$-functions implies that a discrete remnant 
of translation invariance survives compactification:  
KK number conservation.

To obtain 4-d chiral fermions, UED is compactified on an orbifold,
and this introduces fixed points on which interactions that
break KK number conservation could exist.  Generically, KK number
conservation is broken to a subgroup called KK parity 
\cite{Appelquist:2000nn} by brane-localized interactions 
that can arise radiatively \cite{Georgi:2000ks}.
The size of the one-loop brane-localized corrections for UED
have been explicitly calculated in Ref.~\cite{Cheng:2002iz},
which we'll discuss more below.
Nevertheless, KK parity remains unbroken so long as no explicit KK parity
violating interactions are added to the orbifold fixed points.
In other words, KK parity is technically natural, in that the
symmetry structure is enhanced when coefficients of these bare 
would-be KK parity violating interactions are taken to zero.  
This is entirely analogous to $R$-parity in supersymmetric models.

KK parity can be written succinctly as $P_{\rm KK} = (-1)^k$ 
for the $k^{\rm th}$ KK mode.  This implies:
\begin{itemize}
\item The lightest level-one KK mode is stable.  
\item Odd level KK modes can only be produced in pairs.
\item Direct couplings to even KK modes occur through
      brane-localized, loop-suppressed interactions.
\end{itemize}
We'll now discuss these implications of KK parity on the 
phenomenology of the UED model.

\subsection{Corrections to Electroweak Precision Observables}

The typical problem with additional gauge bosons that
couple to light fermions is that they can give large contributions
to electroweak precision observables.  Consider the quintessential
observable, the $Z$-width.  Given measurements of $M_Z$, $G_F$, 
and $\alpha_{\rm em}(M_Z)$, the $Z$ width can be calculated at tree-level.
New contributions to the width potentially arise from the exchange
of heavier gauge bosons, but such contributions do \emph{not} exist 
in UED models since KK parity forbids tree-level couplings of 
$Z^{(k)}$ with the fermion zero modes as well as the 
$Z^{(k)} Z^{(0)} H^{(0)} {H^{(0)}}^\dag$ four-point coupling.

Through one-loop interactions, however, there are calculable
corrections\footnote{Distinguished from cutoff scale contributions, 
discussed below.} to the electroweak precision observables.  
Using the parameterization given by Peskin and Takeuchi
\cite{Peskin:1991sw},
the contributions to $S$ and $T$ 
were calculated in Ref.~\cite{Appelquist:2000nn}.  
They found
\begin{eqnarray}
T = \sum_{j} D_j \left( T_j^t + T_j^h + T_j^V \right)
\end{eqnarray}
where the sum is over all modes up to the cutoff scale of the
$D$-dimensional theory, and $D_j$ is the density of states
at each level $j$.  The individual contributions are
\begin{eqnarray}
\alpha T_j^V &=& -\frac{\alpha}{4 \pi \cos^2 \theta_W} 
                 \frac{(2 n + 1) M_W^2}{6 M_j^2} \\
\alpha T_j^h &=& -\frac{\alpha}{4 \pi \cos^2 \theta_W}
                 \frac{5 m_h^2 + 7 M_W^2}{12 M_j^2} \\
\alpha T_j^t &\simeq& \frac{m_t^4}{8 \pi^2 v^2 M_j^2} 
\end{eqnarray}
where $M_j = j/R$ is $j^{\rm th}$ Kaluza-Klein mass level.
Using the experimental values for the SM parameters, 
the $T$ parameter is roughly
\begin{equation}
T \simeq 0.76 \sum_{j} D_j \frac{m_t^2}{M_j^2} \; .
\end{equation}
A similar calculation can be done for $S$,
\begin{equation}
S \simeq 0.01 \sum_j D_j \frac{m_t^2}{M_j^2}
\end{equation}
where the contribution to the isospin-breaking parameter $T$
is two orders of magnitude larger than the isospin-preserving
parameter $S$.  There is no large contribution to $S$ because
the heavy KK quarks acquire their mass dominantly from the vector-like 
contribution arising from compactification.
Note that the sum over states for these electroweak parameters is
\begin{eqnarray}
\mbox{convergent}      &                        & D=5 \nonumber \\
\mbox{log divergent}   & \quad \mbox{for} \quad & D=6 \nonumber \\
\mbox{power divergent} &                        & D>6 \; . \nonumber
\end{eqnarray}
Contemplating UED models with $D>6$ therefore appear somewhat
problematic, since even the calculable contribution to the
electroweak precision observables diverges.  

In any case, using the calculable corrections we can find a 
lower bound on the inverse radius of the extra dimensions.  
Ref.~\cite{Appelquist:2000nn} required the moderately loose
constraint $T \lsim 0.4$ which leads to
\begin{eqnarray}
\frac{1}{R} &\gsim& \left\{ \begin{array}{lcl}
300 \; \mbox{GeV} & \qquad & D=5 \\
500 \; \mbox{GeV} &        & D=6 \; . \end{array} \right.
\end{eqnarray}
These bounds are probably a bit too low given the latest
electroweak fits \cite{Eidelman:2004wy}, but in any case the
bound is in the several hundred GeV range. 
These bounds on UED dimensions should be contrasted with
those that result from extra dimensions that are \emph{not}
universal, i.e., SM gauge bosons living in higher dimensions 
with SM fermions localized to 4-d.
For example, Ref.~\cite{Rizzo:1999br} found
constraints on the inverse size of an extra dimension of this type 
in the several TeV range.

\subsection{UED Cutoff Scale}

We have alluded to the fact that UED models are effective 
theories of extra dimensions with a cutoff scale.
What is the cutoff scale?  Since gauge couplings in 
extra dimensional theories are dimensionful, i.e.\
$\alpha_D$ has mass dimension $-n$, a rough guess is 
\begin{equation}
\Lambda \sim \frac{4 \pi}{\alpha_D^{1/n}} 
\end{equation}
where I have been excessively naive about my NDA counting.
Matching this $D$-dimensional gauge coupling to a 4-d coupling
of the SM,
\begin{equation}
\frac{1}{g^2} = \frac{(\pi R)^n}{g_D^2}
\end{equation}
we obtain
\begin{equation}
\Lambda R \sim \frac{4}{\alpha^{1/n}} \sim 
\left\{
\begin{array}{lcl}
30 & \mbox{for} & D=5 \\
10 & \mbox{for} & D=6
\end{array} \right.
\end{equation}
which is roughly what is expected.  A 4-d version of the same
calculation, which is arguably better defined, sums over the
number of KK particles running in loops to determine the scale
of strong coupling.  In this way of counting, since the number
of KK modes is proportional to $n^2$, we would expect
$\Lambda_{6-d} R \simeq \sqrt{\Lambda_{5-d} R}$, and thus
$\Lambda_{6-d} R \simeq 5$.  These are the typical numbers given 
for the cutoff scales of the 5-d and 6-d theories.

Cutoff scales that are only about one order of magnitude
above the compactification scale may be problematic in other
ways.  While proton decay and lepton number violating operators
can be suppressed or eliminated in six dimensions, given a
cutoff only a factor of $5$ above the compactification scale
one ought to be concerned about other higher dimensional operators
that violate flavor symmetries or custodial SU(2).  
Also, the cutoff scale may be even lower than the above estimates 
suggest.  Ref.~\cite{Chivukula:2003kq} found that requiring 
scattering amplitudes satisfy the unitarity bound results in
rather low estimates for the scale where strong coupling appears,
only a small factor above the compactification scale.

\subsection{UED in 5-d:  The Spectrum}

For the remainder of the discussion, I want to focus on 
the spectrum of the 5-d version of UED.  Fortunately, the spectrum 
depends linearly on $1/R$ and only logarithmically on $\Lambda R$,
which we will leave as a free parameter varied in some reasonable range.
We will also ignore the effects of higher dimensional operators 
suppressed by the cutoff scale.  What we will do is to examine
more closely the spectrum of UED and the implications for 
collider searches and for the possibility of having a dark matter
candidate.  I'll briefly mention the ways in which a UED dark 
matter candidate could be detected, emphasizing the difference
from a typical supersymmetric candidate.

The spectrum of the 5-d UED model consists of all of the particles
of the SM and their Kaluza-Klein excitations.  Let's focus on
the first KK level.  At tree-level, the masses of the KK particles
are simply
\begin{equation}
m_{KK}^2 = \frac{1}{R^2} + m_{\rm SM}^2
\end{equation}
where $m_{SM}$ is the mass of level-zero (Standard Model) particle.
This suggests a high degree of degeneracy for the KK excitations
of light SM particles, but this degeneracy is not preserved beyond
tree-level.  Indeed, in Ref.~\cite{Cheng:2002ab} it was 
realized that radiative corrections to the KK particle masses
are often much larger than the tree-level SM contribution.

There are two classes of calculable radiative corrections.  
One arises from diagrams involving bulk loops, namely particles 
that traverse around the circle (or actually from one side to
the other, in an orbifold).  The loops are non-contractable,
with finite extent in the extra dimension, implying they give finite 
corrections to the masses.  The second class of corrections
involves brane-localized kinetic terms that appear on the boundaries 
of the orbifold.  These corrections are logarithmically sensitive to
the cutoff scale of the theory.  The generic form of this correction
to the Lagrangian is
\begin{eqnarray}
\delta L &=& \left( \delta(y) + \delta(y - \pi R) \right) \frac{R g^2}{128 \pi}
\ln \frac{\Lambda^2}{\mu^2} \times \nonumber \\
& &{} \left[ \overline{F}_+ i \slashchar{\partial} F_+
             + 5 (\partial_5 \overline{F}_-) F_+
             + 5 (\overline{F}_+ (\partial_5 F_-) \right]
\end{eqnarray}
where $F_+$ and $F_-$ are the components of a bulk 
four-component spinor corresponding to any of the SM fermions.
These corrections are necessarily logarithmically sensitive 
to the cutoff scale.

\begin{figure}[t]
\centerline{\includegraphics[width=1.0\hsize]{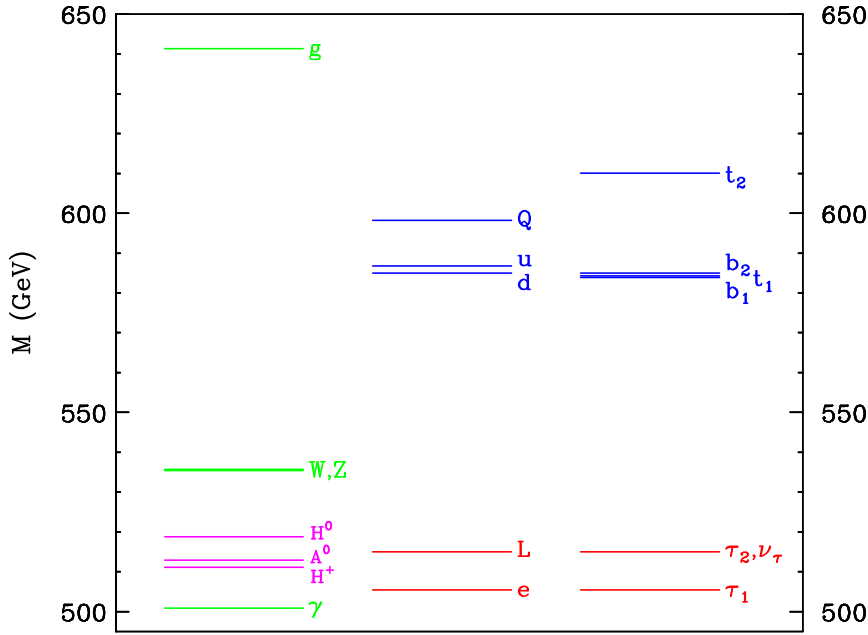}}
\caption{One-loop corrected mass spectrum of
the first KK level for $R^{-1}=500$ GeV, $\Lambda R = 20$
and $m_h=120$ GeV.  The states $t_{1,2}$, $b_{1,2}$, and $\tau_{1,2}$
correspond to the mass eigenstates of the first KK excitations of
the left- and right-handed SM fermions.  
(Fig.~1 from Ref.~\protect\cite{Cheng:2002ab}.)}
\label{UED-spectrum-fig}
\end{figure}

The shifts in the KK masses resulting from these two classes
of radiative corrections are \cite{Cheng:2002iz}:
\begin{eqnarray} 
\delta(m^2_{B^{(n)}}) &=& \frac{g'^2}{16 \pi^2 R^2} 
  \left( \frac{-39}{2} \frac{\zeta(3)}{\pi^2} -\frac{n^2}{3} \ln \, 
  \Lambda R \right) \nonumber \\
\delta(m^2_{W^{(n)}}) &=& \frac{g^2}{16 \pi^2 R^2} \left ( 
  \frac{-5}{2} \frac{\zeta(3)}{\pi^2} + 15 n^2 \ln \,
  \Lambda R \right )\nonumber \\
\delta(m^2_{g^{(n)}}) &=& \frac{ g_3^2}{16 \pi^2 R^2} \left ( 
  \frac{-3}{2} \frac{\zeta(3)}{\pi^2} + 23 n^2 \ln \,
  \Lambda R \right )\nonumber \\
\delta(m_{Q^{(n)}}) &=& \frac{n}{16 \pi^2 R} \left ( 6 g_3^2+ \frac{27}{8} 
g^2 + \frac{1}{8} g'^2 \right) \ln \, \Lambda R \nonumber \\
\delta(m_{u^{(n)}}) &=& \frac{n}{16 \pi^2 R} \left ( 6 g_3^2+  
 2 g'^2 \right) \ln \, \Lambda R \nonumber \\
\delta(m_{d^{(n)}}) &=& \frac{n}{16 \pi^2 R} \left ( 6 g_3^2+  \frac{1}{2} 
g'^2 \right) \ln \, \Lambda R \nonumber \\
\delta(m_{L^{(n)}}) &=& \frac{n}{16 \pi^2 R} \left ( \frac{27}{8} 
g^2 + \frac{9}{8} g'^2 \right) \ln \, \Lambda R \nonumber \\
\delta(m_{e^{(n)}}) &=& \frac{n}{16 \pi^2 R}\frac{9}{2} g'^2 \ln \, \Lambda R 
\; . \label{MassCorr}
\end{eqnarray}
All of the non-colored KK excitation masses are within about 10\% 
of $m_{\gamma^{(1)}}$ up to moderately high values of the cutoff scale 
($\Lambda R \approx 30$).  The strongly interacting particles
are somewhat heavier, up to perhaps $20\%-25\%$ above $1/R$.
Fig.~\ref{UED-spectrum-fig} gives an example of this spectrum
for a compactification size relevant to upcoming collider
experiments.

I should emphasize that there are several assumptions built into
this radiatively-corrected spectrum.  One is that the matching 
contributions to the brane-localized kinetic terms are assumed to be 
zero when evaluated at the cutoff scale.  This leads to a finite 
correction that should be compared against a log-enhanced 
correction.  However, since the log is relatively small, 
of order $\ln 30 \sim 3.4$, the finite contribution 
could easily compete or dominate over this correction.
Also, the spectrum assumes that there are no brane-localized 
quadratically-divergent contributions to the Higgs (mass)$^2$.

Nevertheless, it is intriguing that the spectrum is so qualitatively
similar to a moderately degenerate supersymmetric spectrum.
The spin of the KK excitations is of course equal to the spin
of the corresponding SM (zero mode) field, whereas superpartners 
have spin that differ by $1/2$ from their SM counterparts.  
Unfortunately, measurements of the spin of newly discovered 
heavy particles at hadron colliders is not easy.  This had led 
to suggestions that a KK spectrum could easily be mistaken for 
a degenerate supersymmetric spectrum \cite{Cheng:2002ab}.  

Another similarity to supersymmetry is that UED possesses
an auxiliary discrete symmetry that (if exact) forces
pair production of the lightest level-one KK excitations and
prevents the lightest level-one KK excitation from decaying
into SM particles.  The latter property implies that a stable
particle exists in the spectrum, potentially a dark matter candidate.

\begin{figure}[t]
\centerline{\includegraphics[width=1.0\hsize]{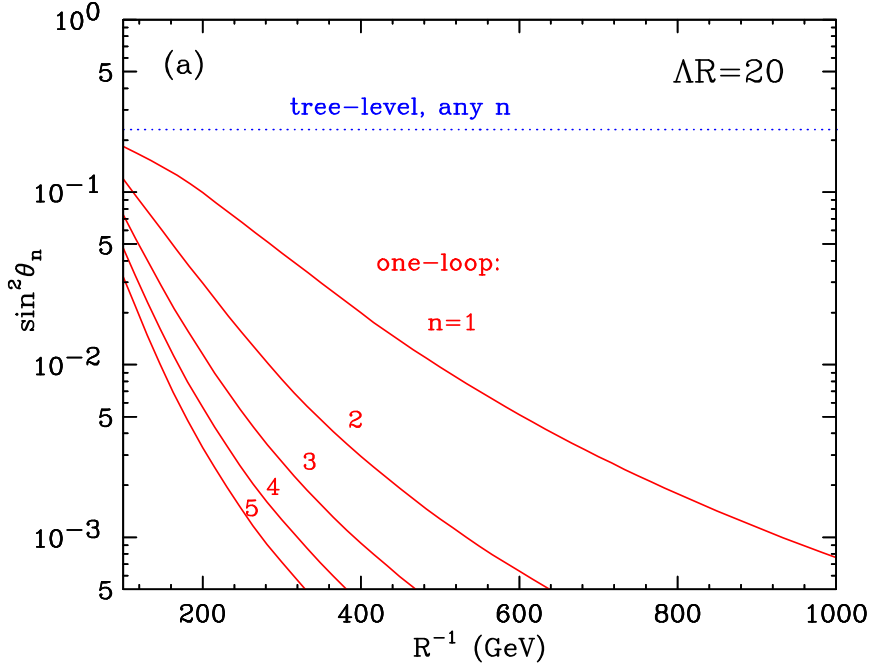}}
\caption{The effective Weinberg angle $\theta^{(n)}$ 
that determines the gauge content of the lightest level-$n$ KK mode,
$\gamma^{(n)} = \cos \theta^{(n)} B^{(n)} - \sin \theta^{(n)} W^{3(n)}$.
(Fig.~5 from Ref.~\protect\cite{Cheng:2002iz}.)}
\label{Weinberg-angle-fig}
\end{figure}

If the spectrum of the level-one KK excitations follows precisely
that of Eqs.~(\ref{MassCorr}), then the lightest KK particle is 
the KK photon.
Saying ``KK photon'' is somewhat misleading, however, since the
neutral gauge boson mass matrix in the ($B^{(n)}$, $W^{3(n)}$) basis
\begin{equation}
\left( \begin{array}{cc} 
\frac{n^2}{R^2} + \delta m^2_{B^{(n)}} + \frac{1}{4} g'^2 v^2 
  & \frac{1}{4} g'g v^2 \\
\frac{1}{4} g'g v^2 
  & \frac{n^2}{R^2} + \delta m^2_{W^{(n)}} + \frac{1}{4} g^2 v^2 
\end{array} \right)
\end{equation}
depends on both the tree-level contributions (proportional to $v^2$)
and the radiative corrections.  The effective mixing angle,
$\sin^2 \theta_W^{(n)}$ for the $n^{\rm th}$ mode is much smaller
than the Weinberg angle, shown in Fig.~\ref{Weinberg-angle-fig}.
Clearly, for $1/R \gsim 500$ GeV combined with $\Lambda R \sim 20$,
$\gamma^{(1)} \simeq B^{(1)}$ to within a percent, and so for all
subsequent purposes we can consider them equivalent.  
From now on I'll just use $B^{(1)}$ to refer to the lightest
level-one KK excitation of the neutral gauge bosons.

\subsection{UED Dark Matter}

It is amusing that $B^{(1)}$ happens to be the candidate for dark
matter in UED models.  The close analogy with supersymmetry would seem 
to continue here, since the supersymmetric partner to the hypercharge
gauge boson, the Bino ($\tilde{B}$), is the typical candidate for
supersymmetric dark matter.  But, this is where the similarity ends.

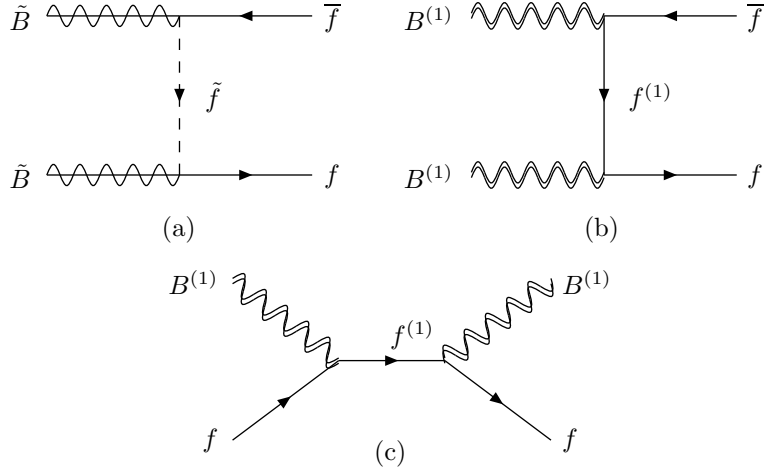
\begin{figure}[t]
\begin{picture}(300,200)
  \Text( 14, 180 )[r]{$\tilde{B}$}
  \Text( 14, 120 )[r]{$\tilde{B}$}
  \Photon( 20 , 180 )( 70 , 180 ){4}{5}
  \Line(   20 , 180 )( 70 , 180 )
  \Photon( 20 , 120 )( 70 , 120 ){4}{5}
  \Line( 20 , 120 )( 70 , 120 )
  \ArrowLine( 120, 180 )( 70, 180 )
  \DashArrowLine( 70, 180 )( 70, 120 ){4}
  \ArrowLine( 70, 120 )( 120, 120 )
  \Text( 80 , 150 )[l]{$\tilde{f}$}
  \Text( 125 , 180 )[l]{$\overline{f}$}
  \Text( 125 , 120 )[l]{$f$}
  \Text( 70 , 100 )[c]{(a)}
  \Text( 174, 180 )[r]{$B^{(1)}$}
  \Text( 174, 120 )[r]{$B^{(1)}$}
  \Photon( 180 , 181 )( 230 , 181 ){4}{5}
  \Photon( 180 , 179 )( 230 , 179 ){4}{5}
  \Photon( 180 , 121 )( 230 , 121 ){4}{5}
  \Photon( 180 , 119 )( 230 , 119 ){4}{5}
  \ArrowLine( 280, 180 )( 230, 180 )
  \ArrowLine( 230, 180 )( 230, 120 )
  \ArrowLine( 230, 120 )( 280, 120 )
  \Text( 240 , 150 )[l]{$f^{(1)}$}
  \Text( 285 , 180 )[l]{$\overline{f}$}
  \Text( 285 , 120 )[l]{$f$}
  \Text( 230 , 100 )[c]{(b)}
  \Text( 85, 80 )[r]{$B^{(1)}$}
  \Text( 85, 20 )[r]{$f$}
  \Photon(  90 , 81 )( 130 , 51 ){4}{5}
  \Photon(  90 , 79 )( 130 , 49 ){4}{5}
  \Photon( 170 , 51 )( 210 , 81 ){4}{5}
  \Photon( 170 , 49 )( 210 , 79 ){4}{5}
  \ArrowLine(  90, 20 )( 130, 50 )
  \ArrowLine( 130, 50 )( 170, 50 )
  \ArrowLine( 170, 50 )( 210, 20 )
  \Text( 150 , 60 )[l]{$f^{(1)}$}
  \Text( 215 , 80 )[l]{$B^{(1)}$}
  \Text( 215 , 20 )[l]{$f$}
  \Text( 150 , 15 )[c]{(c)}
\end{picture}
\caption{Relevant annihilation and scattering processes 
for (a) supersymmetry and (b),(c) UED.  The supersymmetric 
annihilation diagram (a) is $s$-wave suppressed by a factor 
$m_f^2/m_{B^{(1)}}$, whereas the UED diagram (b) is unsuppressed.
Observable annihilation in the Sun occurs through diagram (b) with $f = \nu$.
Annihilation in the galactic neighborhood to positrons occurs through 
diagram (b) with $f = \ell$.  Scattering off nuclei occurs via
diagram (c) with $f = q$, suitably ``dressed'' into a 
proton or neutron.}
\label{annihilation-fig}
\end{figure}

In supersymmetry, Bino annihilation typically proceeds through
sfermion exchange, shown in Fig.~\ref{annihilation-fig}(a).  
Since the Binos
are Majorana fermions, Fermi statistics requires that they have
their spins oppositely directed when prepared in an initial
$s$-wave.  This means that a chirality flip of the fermions 
is required, and thus a mass insertion in the diagram.  
This causes the cross section 
to be suppressed by a factor $m_f^2/m_{\tilde{B}}^2$.

\begin{figure}[t]
\centerline{\includegraphics[width=1.0\hsize]{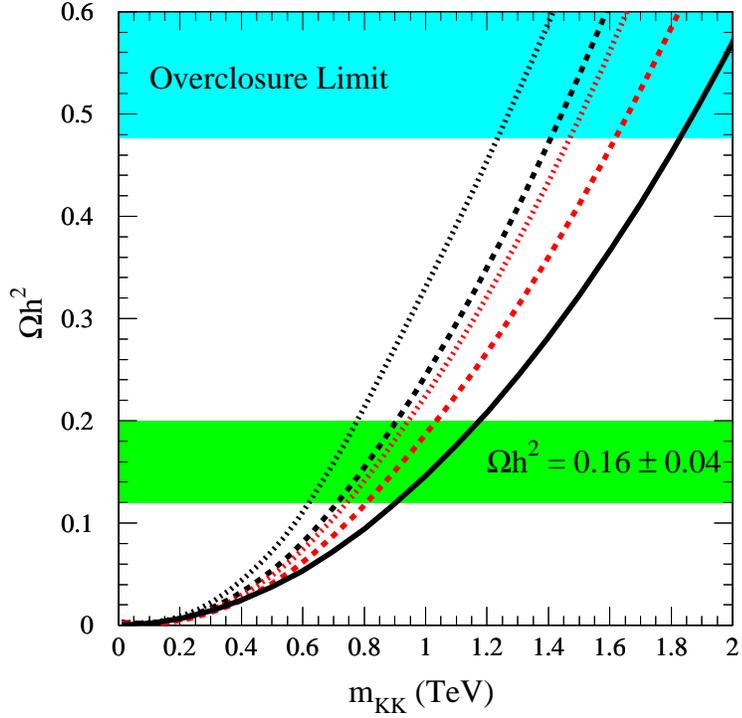}}
\caption{Prediction for $\Omega_{B^{(1)}} h^2$.
The solid line is the case for $B^{(1)}$ alone, 
and the dashed and dotted lines correspond
to the case in which there are one (three) flavors of nearly degenerate 
$e_R^{(1)}$.  For each case, the black curves (upper of each pair) 
denote the case where the fractional mass difference between 
the RH KK lepton and $B^{(1)}$ is 1\%, while the red curves 
(lower of each pair) correspond to 5\%.
Note that the ``favorable range'' of $\Omega h^2$ for dark matter 
is now out-of-date; current cosmological data suggest 
$\Omega h^2 \sim 0.1$, just below the bottom of the shaded band.  
(Fig.~3 from Ref.~\protect\cite{Servant:2002aq}.)}
\label{ST-fig}
\end{figure}

In UED, $B^{(1)}$ annihilation also proceeds through KK fermion
annihilation shown in Fig.~\ref{annihilation-fig}(b), 
but because the incoming states 
are bosons, there is no $s$-wave suppression.  This means that
the mass range for $B^{(1)}$ to make up the dark matter of the Universe
is significantly higher than the range of Bino masses for 
supersymmetric dark matter.
We can estimate the thermally-averaged annihilation cross section
by assuming that only diagrams of type (b) shown in 
Fig.~\ref{annihilation-fig}
are present.  Given the KK spectrum above, Eqs.~(\ref{MassCorr}),
the radiative correction to level-one KK fermions is typically
at the few to tens of percent level.  A reasonable approximation
is to assume that the mass of the exchanged KK particle $f^{(1)}$ 
is roughly degenerate with $B^{(1)}$.  The cross section is then
simply $1/m_{B^{(1)}}^2$ times the coupling factors, which are
just $g_1^4 \sum_i (Y_f^{i})^4$.  The final result is
\begin{equation}
\langle \sigma v \rangle = \frac{95 g_1^4}{324 \pi m_{B^{(1)}}^2}
\end{equation}
where this cross section with numerical factors was  
worked out in Ref.~\cite{Servant:2002aq}.  If you take a 
rough estimate of the relic density given in Kolb and Turner 
\cite{Kolb:1990vq}
\begin{equation}
\Omega h^2 \sim 
   \frac{0.77 \times 10^{-37} \; {\rm cm}^2}{\langle \sigma v \rangle}
\end{equation}
and then plug in the thermally-averaged cross section above, 
one obtains
\begin{equation}
\Omega h^2 \sim 0.1 \left( \frac{m_{B^{(1)}}}{1 \; {\rm TeV}} \right)^2
\end{equation}
which is accurate to within about 15\% of the numerical results given 
in Ref.~\cite{Servant:2002aq}.  A plot of the relic density 
is shown in Fig.~\ref{ST-fig}, along with several additional
curves that represent including coannihilation with one to
three generations of the level-one KK excitations of the
right-handed leptons.  

An interesting outcome of the analysis of Ref.~\cite{Servant:2002aq} 
is that coannihilation\footnote{See 
Ref.~\cite{Griest:1990kh} for a nice general discussion of 
the effects of coannihilation.}
with light KK leptons causes an \emph{increase}
in the effective annihilation cross section and thus a 
\emph{decrease} in the mass range of the KK particle.
This happens because the additional KK particles, when
close enough in mass to $B^{(1)}$, have small annihilation
and coannihilation cross sections, freeze 
out later, causing them to act as additional components
to dark matter.  These right-handed (RH) KK leptons decay into $B^{(1)}$
after their mutual interactions are too slow compared with
the expansion rate, and thus they decay into $B^{(1)}$,
boosting the $B^{(1)}$ relic density, or equivalently lowering
the mass range of $B^{(1)}$ when the relic density is held fixed.

This result, however, is unique to right-handed KK leptons.
As shown by two groups Refs.~\cite{Burnell:2005hm,Kong:2005hn},
coannihilation with left-handed KK leptons, KK quarks, 
KK gluons, etc., all have cross sections that are larger
than $B^{(1)}$ annihilation, causing the total effective
cross section to go up.  Holding the relic density fixed,
this implies the mass range of $B^{(1)}$ must also increase.
If the KK quarks and KK gluon are below about $1.1$ times the 
mass of $B^{(1)}$, these coannihilation effects can cause
the mass range for $B^{(1)}$ to go up to the few TeV range.
On face value, such a small separation between the mass of $B^{(1)}$ 
and the strongly interacting level-one KK particles is not expected 
from the radiative corrections to the masses of the first KK level 
computed in \cite{Cheng:2002iz}.  However, if the cutoff scale is not much
larger than the compactification scale, and thus matching corrections 
are comparable in size while opposite in sign to compensate, 
the level-one KK spectrum could be much more degenerate.

UED dark matter can be detected by the usual methods,
namely direct detection by scattering off nuclei, 
as well as indirect detection through annihilation 
in the Sun to neutrinos or annihilation in our galactic
neighborhood to positrons.  

\subsection{Direct Detection of UED Dark Matter}

First consider the direct detection of $B^{(1)}$ dark matter.  Dark
matter particles are currently non-relativistic, with velocity $v \sim
10^{-3}$.  For weak scale dark matter, the recoil energy from
scattering off nuclei is far less than for scattering off electrons,
and thus one need only consider elastic scattering off nucleons
and nuclei.

At the quark level, $B^{(1)}$ scattering goes through KK quarks,
such as shown in Fig.~\ref{annihilation-fig}(c).  The amplitudes
and cross sections for the quark level processes are easy to 
calculate, but then these processes must be convoluted 
with structure functions for nucleons and nuclei.
The interactions divide into spin-dependent and spin-independent
parts \cite{Goodman:1984dc}.  Higgs exchange contributes to scalar
couplings, while KK quark exchange contributes to both.

\begin{figure}[t]
\centerline{\includegraphics[width=1.0\hsize]{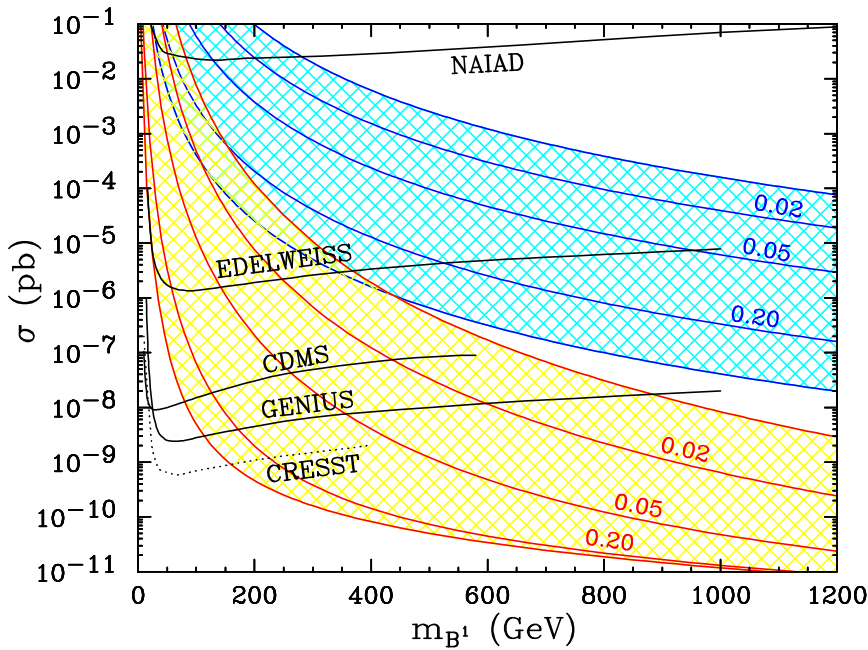}}
\caption{
Predicted spin-dependent proton cross sections (dark-shaded, blue),
along with the projected sensitivity of a 100 kg NAIAD
array; and predicted spin-independent proton cross
sections (light-shaded, red), along with the current EDELWEISS
sensitivity, and projected sensitivities of
CDMS, GENIUS, and CRESST.  
The predictions are for $m_h = 120$ GeV and $0.01 \le r =
(m_{q^1} - m_{B^{(1)}}) / m_{B^{(1)}} \le 0.5$, 
with contours for specific intermediate $r$ labeled.
(Fig.~1 from Ref.~\protect\cite{Cheng:2002ej}.)}
\label{KK-direct-fig}
\end{figure}

In Refs.~\cite{Cheng:2002ej,Servant:2002hb}
both spin-independent and spin-dependent cross sections 
were calculated and are shown in Fig.~\ref{KK-direct-fig}.  
This figure assumes all level-one KK quarks are
degenerate with mass $m_{q^{(1)}}$ that is different from $m_{B^{(1)}}$.
Projected sensitivities of near future experiments are also shown in
Fig.~\ref{KK-direct-fig}.  For scattering off individual nucleons, scalar
cross sections are suppressed relative to spin-dependent ones by $\sim
m_p/m_{B^{(1)}}$.  However, this effect is compensated in large nuclei 
where spin-independent rates are enhanced by $\sim A^2$
($A$ is the nuclei mass number).
In the case of
bosonic UED dark matter, the latter effect dominates, and the
spin-independent experiments have the best prospects for detection
with sensitivity to $m_{B^{(1)}}$ far above current limits.

\subsection{Indirect Detection of UED Dark Matter}

\begin{figure}[t]
\centerline{\includegraphics[width=0.8\hsize]{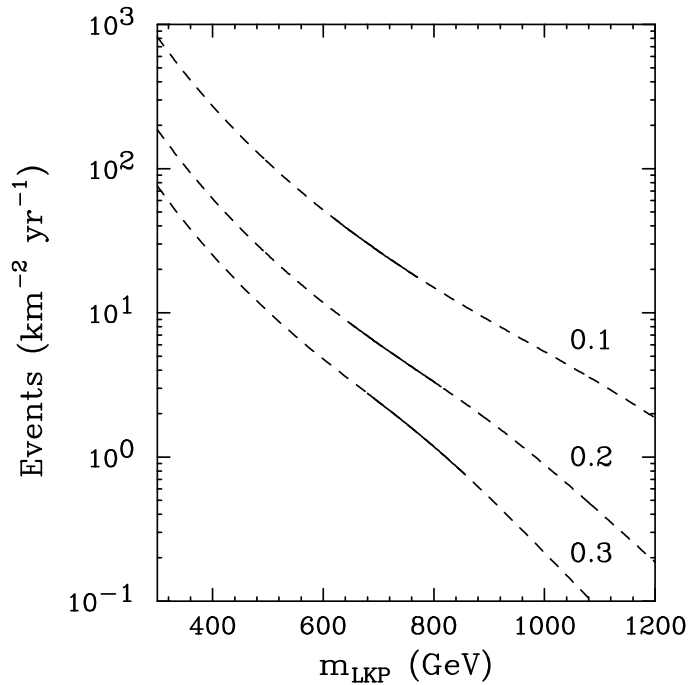}}
\caption{The number of events per year in a detector with effective 
area equal to one square kilometer.  Contours are shown for 
$r_{q^{(1)}_R}=0.1$, $0.2$, and $0.3$.  The $r_{q^{(1)}_R}=0.3$
is shown merely for comparison, since this mass ratio is 
larger than would be expected from the one-loop radiative correction 
calculations of the KK mode masses.
The relic density of the $B^{(1)}$'s lies within the range
$\Omega_{B^{(1)}} h^2 = 0.16 \pm 0.04$ for the solid sections 
of each line.  Matching to the WMAP data, in which 
$\Omega_{B^{(1)}} h^2 \sim 0.1$ is preferred, corresponds roughly
to the left-hand side dash-to-solid transition for each curve.
(Fig.~2 from \protect\cite{Hooper:2002gs}.)}
\label{KK-sun-fig}
\end{figure}

Weakly interacting dark matter particles are expected to become 
gravitationally trapped in large bodies, such as the Sun, and 
annihilate into neutrinos or other particles that decay into neutrinos.
The calculation of the flux of neutrinos from particle dark matter 
annihilation in the
Sun has been explored in some detail, particularly the case
of neutralino dark matter (for reviews, see 
Refs.~\cite{Jungman:1995df,Bertone:2004pz}).
The basic idea is to 
begin with the relatively well-known local dark matter density from
the galactic rotation data, compute the interaction cross section 
of the particle dark matter with nuclei in the Sun, compare the capture 
rate with the annihilation rate to determine if these processes
are in equilibrium, and then compute the flux of neutrinos that
result from this rate capture and annihilation.
There is a huge detector at the south pole that has instrumented
a large area of antarctic ice by stringing detectors down deep holes.
The first version of this experiment had an effective area of 
$0.1 \; {\rm km}^2$ (called AMANDA) that is now in progress 
towards expansion to $1 \; {\rm km}^2$ (called IceCube).

The calculation of the annihilation rate in the Sun involves
similar scattering processes to what we found for direct detection,
except that now the dominant process is simply scattering
off protons.  The new ingredient is to determine when the
capture rate equilibrates with the annihilation rate, 
which is determined by the mass of the (core of the) Sun, 
the dark matter density and (relative) velocity, 
as well as the microscopic scattering cross section.  
Since the particle dark matter candidate in UED is rather heavy,
it is not surprising that the elastic scattering is not
particularly large, and a detailed calculation \cite{Hooper:2002gs}
shows that $B^{(1)}$ dark matter just barely comes into
equilibrium after 4.5 billion years.  This gives the
maximal neutrino signal emitted from the Sun.

The actual outgoing flux depends on
the annihilation fraction directly into neutrinos, as well as 
indirectly through decays.  Muon neutrinos are the main actors, 
since at the energies relevant to $\B$ annihilation, 
neutrino telescopes only observe muon tracks generated in 
charged-current interactions.  In Ref.~\cite{Hooper:2002gs}
the heavier level-one KK modes were approximated to have about the
same mass, but this mass was taken to be slightly larger than 
$m_{B^{(1)}}$ by a fraction 
\begin{equation}
r_{f^{(1)}} \equiv \frac{m_{f^{(1)}} - m_{B^{(1)}}}{m_{B^{(1)}}} \; ,
\end{equation}
typically about $0.1$--$0.2$ given the spectrum from Eqs.~(\ref{MassCorr}).  
Using the neutrino energy spectrum, the event rate expected at an 
existing or future neutrino telescope can be calculated.  
This is shown in Fig.~\ref{KK-sun-fig} 
for a detector with an effective area of $1 \; {\rm km}^2$.
Combining the spectrum determined by the one-loop radiative corrections 
with a relic density appropriate for dark matter, the expectation
is to get between a few to tens of events per year at the 
IceCube detector.  

Finally, there are speculations that UED dark matter annihilation
in the galactic halo might account for the positron excess,
see Ref.~\cite{Hooper:2004xn} for details.

\section{Conclusions}

In these lectures I have showed how the phenomenology of 
extra dimensions is very rich, should Nature choose to follow 
one or more of the ideas discussed in this review.  I have tried to 
give a overview of what I perceive to be the main characteristic signals 
of the specific extra dimensional models that I discussed.  
Nevertheless, there are several related models and a host of other 
aspects to extra dimensions that I did not have the time or 
space to review.  This remains a very active field of investigation 
with new ideas continually being developed.  

How likely is any given extra dimensional proposal?
This is not a question that has any scientific answer, even 
though physicists try hard to quantify their qualitative
instincts.  This much can be said with relative certainty:
In all cases the cutoff scale of the Standard Model is 
drastically lowered from the Planck scale to near the 
TeV scale.  Since the lore of quantum gravity is that all
global symmetries are broken by Planck scale effects,
naturalness suggests cutoff-scale suppressed higher dimensional 
operators should appear at the 1/TeV level with order one 
coefficients.  If this were true, all of these models would
be ruled out immediately by $B$ and $L$ violating operators, 
operators leading to FCNC, and operators modifying precision
electroweak observables.  As model builders, we must conclude that
either the cutoff scale (i.e., the quantum gravity scale) is larger 
or there are mechanisms to suppress or eliminate these dangerous effects.
Some of these remarkably creative mechanisms were discussed or 
referenced in the preceding sections.  As phenomenologists, 
however, we are blissfully free to assume that these operators
are suppressed by an unspecified mechanism or simply tuned to
be small, and then we have the opportunity to probe the physics
of these scenarios directly in colliders and indirectly in 
all sorts of ways from astrophysics to table-top experiments.

During the lectures, I repeatedly emphasized that while extra
dimensions are interesting in themselves, the real take home 
lesson is to understand how to turn ``great model A'' into 
``predictions 1,2,3'' and compare with ``experiments X,Y,Z''.
Extra dimensions provide a fascinating and exciting set of models
to illustrate precisely this important exercise.  As we approach
the next energy frontier with the LHC and future experiments,
I hope these lectures instill some of the techniques to turn new physics 
ideas into a testable theory.  Good luck developing your own great ideas
and turning them into calculable phenomenology!

\section*{Acknowledgments}

I thank John Terning, Carlos Wagner, and Dieter Zeppenfeld, 
the organizers of the 2004 ``Physics in $D \ge 4$'' TASI, for the invitation 
to present these lectures and for putting together a fabulous program. 
I particularly thank John Terning for persistent nagging emails 
which ensured the writeup of these lectures was finally completed.
I am grateful to K.~T. Mahanthappa for his hospitality. 
I also thank the many TASI participants for their insightful 
questions and comments, and would particularly like to thank 
Jay Hubisz, Ben Lillie, Patrick Meade, and Brian Murray for 
providing feedback on a written version of these lectures.
This work was supported in part by the Department of Energy 
grant number DE-FG02-96ER40969.



\end{document}